\documentclass{aa}

\usepackage{graphicx}
\usepackage{wasysym}
\usepackage{amssymb}
\usepackage{amsmath}

\usepackage{graphicx} 
\usepackage[latin1]{inputenc} 

\begin{document}

\title{Rotation of rigid Venus : a complete precession-nutation model}

\author{L.~Cottereau \inst{1} \and J.~Souchay\inst{2,1} }


\institute{Observatoire de Paris, Syst\`emes de R\'ef\'erence Temps Espace (SYRTE), CNRS/UMR8630, Paris, France}

\date{}

\abstract
{With the increasing knowledge of the terrestrial planets due to recent space probes it is possible to model their rotation with increasing accuracy. Despite that fact, an accurate determination of Venus precession and nutation is lacking}
{Although Venus rotation has been studied in several aspects, a full and precise analytical model of its precession-nutation motion remains to be constructed. We propose to determine this motion with up-to-date physical parameters of the planet}
{We adopt a theoritical framework already used for a precise precession-nutation model of the Earth, based on a Hamiltonian formulation, canonical equations and an accurate development of the perturbing function due to the Sun. }
{After integrating the disturbing function and applying the canonical equations, we can evaluate the precession constant $\dot{\Psi}$ and the coefficients of nutation, both in longitude and in obliquity. We get $\dot{\Psi}=4474".35/Jcy \pm 66.5 $, corresponding to a precession period of $28965.10 \pm 437$ years. This result, based on recent estimations of the Venus moment of inertia is significantly different from previous estimations. The largest nutation coefficient in longitude with an argument $2L_{S}$ (where $L_{S}$ is the longitude of the Sun) has a 2"19 amplitude and a 112.35 d period. We show that the coefficients of nutation of Venus due to its triaxiality are of the same order of amplitude as these values due to its dynamical flattening, unlike of the Earth, for which they are negligible. }
{We have constucted a complete theory of the rotation of a rigid body applied to Venus, with up-to-date determinations of its physical and rotational parameters. This allowed us to set up a new and better constrained value of the Venus precession constant and to calculate its  nutation coefficients for the first time.}

\maketitle

\section{Introduction}

Among the planets of the Solar system, Venus shows peculiar characteristics : its rotation is retrograde and  very slow. Since its 243.02 d period was determined by radar measurements (Golstein 1964 ,Carpenter 1964), several authors have attempted to explain these two characteristics.  Goldreich and Peale (1970) showed that thermally driven atmospheric tidal torques and energy dissipation  at the boundary between a rigid mantle and a differentially rotating liquid core  are possible mechanisms to maintaine the retrograd spin. Lago and Cazenave (1979) studied the past evolution of the rotation of Venus using the hypothesis that only solar tidal torques and core-mantle coupling have been active since its formation. They found it conceivable that Venus originally had a rotation similar to the other planets and evolved for $4.5\times 10^{9}$ years from a rapid and direct rotation to the present slow retrograde one. Others authors proposed different scenarios by supposing a high value of the initial obliquity (Dobrovoskis,1980, Mc Cue and Dormand,1993). Yoder (1995) gave a full account of the various internal mechanisms acting on Venus'obliquity, such as core friction, CMB (core-mantle boundary) ellipticity, and resonant excitations. Correia and Laskar (2001,2003 and Correia et al., 2003), explored a large variety  of initial conditions in order to cover  possible formation and evolutionary scenarios. They confirmed that despite the variations in the models, only three of the four final spin states of Venus are possible and that the present observed retrograd spin state can be attained by two different processes : one prograde and the other retrograde. Although these various studies concern the evolution of Venus rotation on very long times scales, few attempts have been made to accurately model the rotation for short times scales. An analytical study of the rotation of a rigid Venus model carried out by Habibullin (1995) with rather uncommun parametrization, showing that Venus rotates almost  uniformly  with negligible libration harmonics. In the following, we propose an alternative construction of a rigid Venus rotation model, contradicting to some extent these last results. First we show the characteristics of Venus with respect to the corresponding ones for the Earth (section \ref{2}). We explain the parametrization of the rotation of a rigid Venus using the Andoyer variables (section \ref{3}). We then determine the reference points and planes to study the rotational dynamics of Venus (section \ref{4}). We give the equations of  motion and the analytical developments that allow us to obtain the precession and the nutation of the planet (sections \ref{5}, \ref{6} and \ref{7}). Finally we give the precession, the tables of nutation and the polar motion of Venus (section \ref{8} and \ref{9}). We will discuss our results in  section \ref{10}. In this study we make some approximations : we suppose that the relative angular distance between the three poles (the poles of angular momentum, of figure and of rotation) is very small as is the case for the Earth. Moreover we solve a simplified system where the second order of the potential does not appear. Last, we suppose that the small difference between the mean longitude and the mean anomaly of the Sun which corresponds to the slow motion of the perihelion, can be considered as constant. In sections \ref{3} to \ref{10}, the orbital elements of Venus and their time dependence are required to carry out our numerical calculations. In these sections we used the mean orbital elements given by Simon et al.(1994). These elements are valid over 3000 years with a relative accuracy of $10^{-5}$. This prevents our theory being used in long term calculations, our domain of validity being 3000 years.

\section{Venus characteristics}\label{2}

  From the point of view of its physical characteristics, Venus can be considered as similar to sister of the Earth: the two planets have roughly the same size, mass and mean density (see Table \ref{table1}.).

\begin{table}[!h]
\begin{center}
\resizebox{0.8\hsize}{!}{\begin{tabular}[h]{lccc}
\hline\\
  & Earth & Venus  \\
\hline\\
Semi-major axis & 1.000001 U.A &0.723332 U.A  &   \\
\hline\\
Eccentricity & 0.0167 & 0.0068 &  \\
\hline\\
Inclination  & 0 & 3.39 & \\
\hline\\
Ascending node &174.87 & 7667 &  \\
\hline\\
Period of revolution  & 365.25 d & 224.70 d &   \\
\hline\\
Density & 5.51 $\mathtt{g/cm^3}$ & 5.25 $\mathtt{g/cm^3}$ &  \\
\hline\\
Equatorial radius & 6378 km & 6051 km &  \\
\hline\\
Mass &$ 5.97*10^{24}$ kg&$ 4.87*10^{24}$ kg &  \\
\hline\\
Obliquity  & 23.43 & 2.63 &  \\
\hline\\
Period of rotation  &0.997 d & -243.02 d &  \\
\hline\\
 Triaxiality : $\frac{A-B}{4C}$ & $ -5.34*10^{-6}$ & $ -1.66*10^{-6}$ \\\\
\hline\\
 Dyn. flattening : $\frac{2C-A-B}{2C}$ &  $ 3.27*10^{-3}$ & $ 1.31*10^{-5}$ \\\\
\hline\\
$\frac{B-A}{2(2C-A-B)}$ & 0.0016 & 0.12  \\
\end{tabular}
}
\end{center}
\caption{Comparison between the orbital (at J2000.0) and physical parameters of Venus necessary for our study and the corresponding ones for the Earth.}
\label{table1}
\end{table}

Despite this fact, the rotation of the two planets has with h distinctive characteristics. First, the rotation of Venus is retrograde with a period of 243 days. Furthermore the obliquity of Venus is very small (2.63) compared to that of the Earth (23.44).  Last, as we will see in the following, the perturbating function for the rotational motion of the planets depends on the dynamical flattening $H_{V}=\frac{2C-A-B}{2C}$ and on the triaxiality $T_{V}=\frac{A-B}{4C}$. The Earth has a fast rotation, therefore it has a relatively large dynamical flattening. Moreover its main moments of inertia $A$ and $B$ according to the X and Y axes can be considered equal when comparing them to $C$. Thus the coefficient  of triaxiality is very small with respect to the coefficient of dynamical flattening. On the other hand, Venus has slow rotation. Therefore its coefficient of dynamical flattening is less important than that of the Earth   but its coefficient of triaxility is of the same order as the coefficient of dynamical flattening (see  Table.\ref{table1}). For these reasons, it is important to account for the effect of triaxiality on the rotation of Venus and it is interesting to see what difference will emerge with respect to the Earth's rotation. In Table \ref{table1}, for the sake of clarity we have written the orbital parameters at J2000.0. However throughout this paper the orbital parameters are considered as a function of time (containing terms of first and third order in time).

\section{ Description of the motion of rotation and torque-free motion of Venus }\label{3}

\begin{figure}[htbp]
\center
\resizebox{1.\hsize}{!}{\includegraphics{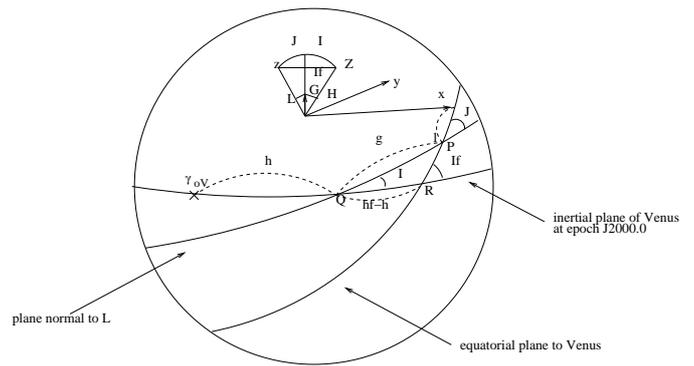}}
 \caption{Relation between the Euler angles and the
Andoyer variables.}
\label{fig1}
\end{figure}

In order to describe in the most complete way the rotation for a rigid model of Venus around its center of mass we need four axes : an inertial axis (arbitrairly chosen) , the figure axis, the angular momentum axis, and the axis of rotation (see Fig.\ref{fig1}.).

$\bullet$ As an inertial plane (0, X, Y) we choose the Venus osculating orbital plane at the epoch J 2000.0. We note $\vernal_{0V}$, the intersection between this orbital plane  and the mean equator of Venus at J2000.0, which is the reference point on the (0, X, Y) plane. The choice of this plane is natural because we want to compute the variation of the obliquity of Venus which is the angle between the orbit of Venus and the plane normal to L.

$\bullet$ The angular momentum axis of Venus is the axis directed along $\overrightarrow{G}$, with $G$ the amplitude of the angular momentum. We denote $h=\vernal_{0V}Q$ and I as respectively the longitude of the node with respect to $\vernal_{0V}$ and the inclination with respect to (0,X,Y).  $g$ and $J$ are the longitude of the node and the obliquity of this same plane with respect to Q and to the equatorial plane .

$\bullet$ The figure axis is the principal axis of Venus directed along Oz and perpendicular to the equatorial plane. We choose it so that it coincides with the axis of the largest moment of inertia $C$. We denote 0x the axis that is pointed toward the origin meridian on Venus, which can be defined in a conventional way. The axis of the figure is determined with respect to the inertial plane (O,X,Y) through the Euler angles  $ h_{f}, I_{f}$. The parameter $\Phi \approx l+g+h$ gives the position of the prime meridian (O,x) with respect to $\vernal_{0V}$.

$\bullet$ The axis of rotation is determined through the variables $h_{r},I_{r}$, i.e. the longitude of the node and the inclination with respect to $\vernal_{0V}$ and the orbit of Venus at J2000.0. For the sake of clarity, these parameters are not represented in Fig.\ref{fig1}.

\vspace{0.5cm}  To describe the motion of the rotation of the rigid Earth, Kinoshita (1977) used Andoyer variables. Therefore we have introduced these variables to apply an analogous theory to Venus. The Andoyer variables (1923) are  (see Fig.\ref{fig1}.) : \\
\begin{itemize}
\item $L$ the component of the angular momentum along the 0z axis
\item $H$ the component of the angular momentum  along the 0Z axis
\item $G$ the amplitude of the angular momentum of Venus
\item $h$ the angle between $\vernal_{0V}$ and the node Q between the orbital plane and the plane normal to the angular momentum.
\item $g$ the angle between the node Q and the node P
\item $l$ the angle between the origin meridian Ox and the node P.
\end{itemize}
\vspace{0.5 cm} This yields:
\begin{equation}\label{papa}
L = G \cdot \cos J \  et \  H = G \cdot \cos I 
\end{equation}
where $I, J$ are respectively the angle between the angular momentum axis and the inertial axis (O,Z), between the angular momentum axis and the figure axis.

By using the spherical trigonometry in the triangle (P, Q, R) (see Fig.\ref{fig1}.), we determine the relations between the variables :
\begin{equation} \label{qelement}
\cos I_{f}=\cos I \cos J -\sin I \sin J \cos g
\end{equation}
\begin{equation}\label{sinus}
\frac{\sin(h_{f}-h)}{\sin J}= \frac{\sin(\Phi-l)}{\sin I}=\frac{\sin g}{\sin I_{f}}.
\end{equation}
The angle $J$ is supposed to be  very small, so by neglecting the second order we obtain (Kinoshita, 1977):
\begin{equation}\label{dodo}
h_{f}= h + \frac{J}{\sin I} \sin g + O(J^2)
\end{equation}
\begin{equation} \label{Jnegl}
I_{f}= I + J\cos g + O(J^2)
\end{equation}
\begin{equation}\label{didi}
\Phi = l + g - J\cot I \sin g + O(J^2).
\end{equation}
$h_{f}$ and $I_{f}$  correspond to the same definition as $h$ and $I$, but for the axis of figure instead of the axis of angular momentum.
The Hamiltonian for the torque-free motion of Venus corresponding to the kinetic energy is :
\begin{equation} \label{ham}
F_{0}= \frac{1}{2}(\frac{\sin ^{2} l}{A}+\frac{\cos ^{2} l}{B})(G^{2}-L^{2})+\frac{1}{2}\frac{L^{2}}{C}
\end{equation}
where $A, B, C$ are the principal moments of inertia of Venus.
 Then the Hamiltonian equations are:
\begin{equation}\nonumber
\frac{d}{dt}(L,G,H)=-\frac{\partial{F_{0}}}{\partial{(l,g,h)}}\nonumber
\end{equation}
\begin{equation}
\frac{d}{dt}(l,g,h)=\frac{\partial{F_{0}}}{\partial{(L,G,H)}}.\nonumber
\end{equation}
The Hamitonian is the full energy of the system. We see the Hamiltonian (\ref{ham}) corresponds to the kinetic energy.
\section{Hamiltonian referred to a moving plane}\label{4}

\begin{figure}[!htbp]
\center
\resizebox{1.\hsize}{!}{\includegraphics{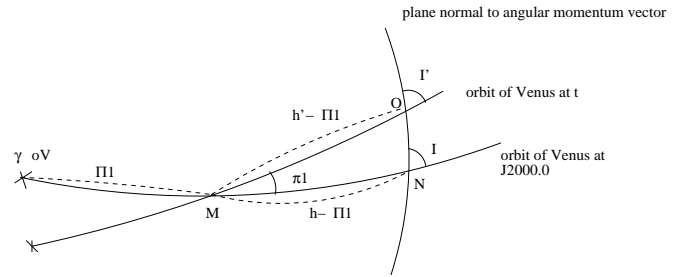}}
 \caption{Motion of the orbit of Venus at the date $t$ with respect to the same orbit at J2000.0. }
\label{PEV}
\end{figure}

This section is based on the pioneering work by Kinoshita (1977) for the rotation of the Earth. As a new reference plane, we choose the plane of the orbit of Venus at the date $t$ instead of the fixed orbital plane at J2000.0. The motion of this new plane is due to the disturbances of the planets and it is defined by two angles $\Pi_{1}$ and $\pi_{1}$ which depend on time (cf Fig.\ref{PEV}.). In subsection \ref{4.3} we will determine accuratly the numerical expressions of $\Pi_{1}$ and $\pi_{1}$.

\subsection{Canonical transformations}\label{tutu}

 We denote $(G', H', L', g', h', l')$ as the new set of Andoyer canonical variables, wich play the same role as $(G, H, L, g, h, l)$ but with respect to the new moving reference plane instead of the inertial one. We denote $I'$ as the angle between the plane normal to angular momentum and the orbit of Venus at $t$. To prove that the transformation is canonical, we  have to determine a complementary function $E$ depending on the new variables :
\begin{equation}\label{an}
G dg+H dh-F_{0}dt = G' dg'+H' dh'-(F_{0}+E)dt.
\end{equation}
Because the tranformation does not depend on $L,l$ these two variables do not appear in (\ref{an}).
In a given spherical triangle (A,B,C) one can show that the following differential relation stands (see appendix) :
\begin{equation} \label{relat}
\mathtt{da= \cos c \cdot db +\cos B \cdot dc+\sin b  \sin C \cdot dA}.
\end{equation}
\begin{figure}[htbp]
\center
\resizebox{0.9\hsize}{!}{\includegraphics{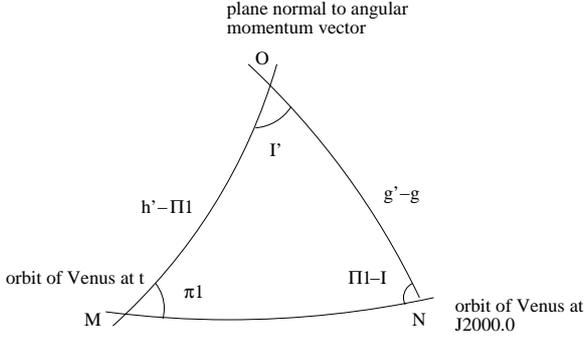}}
 \caption{Details of the spherical triangle M,N,O of Fig.\ref{PEV} which represents the motion of the orbit of Venus at the date t with respect to the same orbit at J2000.0.}
\label{MNNO}
\end{figure}
Applied to the triangle (M,N,O) (see Fig.\ref{MNNO}.) this yields:
\begin{eqnarray}
d(g'-g)&\nonumber=& \cos I' \cdot d(h'-\Pi_{1} ) -\cos I \cdot d(h-\Pi_{1} )\nonumber \\&&+\sin (h'- \Pi_{1}) \sin I' \cdot d\pi_{1}.\nonumber\\
\end{eqnarray}
Using the classical equation on the same triangle :
\begin{equation}
\cos I = \cos I'  \cos \pi_{1}   \\ -\sin I' \sin \pi_{1} \cos ( h'- \Pi_{1}).
\end{equation}
Then, multiplying both sides by $G$ and adding - $F_{o}$ dt, we get :
 \begin{equation}\label{formule canonique}
G \cdot dg + H \cdot dh -F_{o} \cdot dt = G'\cdot dg' + H'\cdot dh'-K  \cdot dt
\end{equation}
where $ G=G' , \ H'=G'\cos I' \ et \ K= F_{0}+E $ \  with
\begin{eqnarray}
E\nonumber&=& H'(1-\cos \pi_{1})\cdot \frac{d\Pi_{1}}{dt}\nonumber \\&& +G'\sin I'\cdot [\frac{d\Pi_{1}}{dt}\cdot \sin \pi_{1} \cos(h'-\Pi_{1})\\ \nonumber&&-\frac{d\pi_{1}}{dt} \cdot \sin(h'-\Pi_{1})].\nonumber\\
\end{eqnarray}
We see that the time does not appear explicitly in $F_{o}$ but appears in the new Hamiltonian $K$ through the variables $\pi_{1}$ and $\Pi_{1}$ expressing the moving orbital plane. Thanks to (\ref{formule canonique}), we have shown that the transformation  from $(G, H, L, g, h, l)$ to $(G', H', L', g', h',l')$ is canonical.This canonical transformation is given in detail in Kinoshita (1977). To calculate the new Hamiltonian, we have to determine the values of $h'-\Pi_{1}$, where  $h'$ is an angle constructed on two different planes, the orbit of Venus at J2000.0 and the orbit of Venus at the date $t$. It is difficult to calculate this angle, so we can do a new transformation, changing $h'$ by $h_{d}$ where $h_{d}$ is the sum of two angles on the orbit of Venus at the date $t$. This small transformation is  particular to the present study and was not done in other works (Kinoshita 1977, Souchay et al. 1999)

\begin{figure}[htbp]
\center
\resizebox{1.\hsize}{!}{\includegraphics{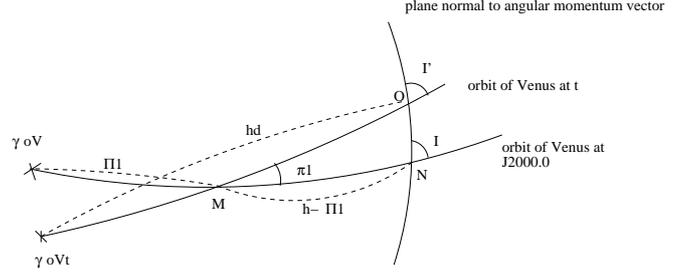}}
 \caption{Motion of the orbit of Venus at the date $t$ with regard to the orbit at J2000.0.}
 \label{PL2}
\end{figure}
 We have $ h_{d}= \overline{\vernal_{0V_{t}}M}+\overline{MO}$ where $ \vernal_{0V_{t}} $ is the so-called "departure point" for Venus, which constitutes our new reference point. The choice of this point is justified by the condition of non-rotation wich characterizes it, described in detail by Guinot (1979) and Capitaine et al (1986). $ \vernal_{0V_{t}} $ is the natural point to measure any motion along the moving plane.
$ h_{d}$ is given by the equation :
 \begin{equation} \label{fifi}
h_{d} = h + s
 \end{equation}
 with
 \begin{equation}
 s = \int(1-\cos\pi_{1}) d\Pi_{1}
 \end{equation}
 To prove that this transformation is canonical, we can determine a function $E'$ depending on the new variables such that :
\begin{equation}\label{gg}
H'dh'-K dt= H'dh_{d}-(K+E')dt.
\end{equation}
Because the tranformation does not depend on $G$ and $g$, these two variables do not appear in (\ref{gg}).
Using equation (\ref{fifi}) we have:
\begin{equation}
H'ds = - E'dt.
\end{equation}
Thus :
\begin{equation}
- H'\frac{ds}{dt} = E'.
\end{equation}
Notice that $s$ is a very small quantity, so that to a first approximation $E'$ is negligible.
Finally, we have a canonical transformation with $K'$ as a new Hamiltonian of the system :
\begin{eqnarray}
K'\nonumber&=&F_{0}+E+E'\nonumber \\&=&F_{0}+G'\sin I'\cdot [\frac{d\Pi_{1}}{dt}\cdot \sin \pi_{1} \cos(h_{d}-\Pi_{1})-\nonumber\\&&-\frac{d\pi_{1}}{dt} \cdot \sin(h_{d}-\Pi_{1})].\nonumber\\
\end{eqnarray}

\subsection{Ecliptic coordinates of the orbital pole and reference point $\vernal{0V}$}

\begin{figure}[htbp]
\center
\resizebox{1.\hsize}{!}{\includegraphics{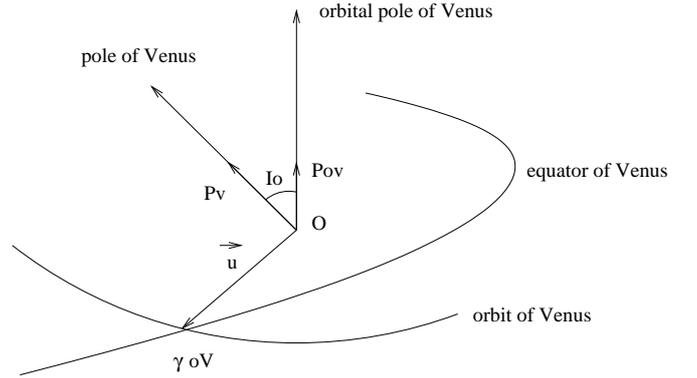}}
 \caption{Relation between the venusian orbital pole  and the venusian pole of angular momentum at the date $t = 0$. }
 \label{OPV}
\end{figure}

To calculate the angles that characterize the moving plane of the orbit of Venus  at the date $t$, a reference point on the orbit of Venus  at J2000.0 is needed. For the Earth the reference point is the vernal equinox $\vernal_{0}$  which coincides  with the ascending node of the ecliptic of J2000.0 with the celestial equator of the same date. In a similar way we determine the coordinates of the ascending node $\vernal_{0V}$ of the orbit of Venus at J2000.0 with respect to the equator of Venus at this same date. To calculate the coordinates of $\vernal_{0V}$ we need $\overrightarrow{P_{ov}}$, the unit vector directed towards the orbital pole at $t=0$ defined by :
\begin{eqnarray}
\overrightarrow{P_{ov}}=\begin{pmatrix}
\sin i_{0} \sin \Omega_{0}\\ -\sin i_{0} \cos \Omega_{0}  \\ \cos i_{0}
\end{pmatrix}
\end{eqnarray}
where $i_{0}$ is the angle between the orbit of Venus and the ecliptic at J2000.0. $\Omega_{0}$ is the longitude of the ascending node at the same date. Simon et al.(1994) give :
\begin{equation}
i_{0}=3.39466189 \quad \Omega_{0} = 76.67992019.
\end{equation}
The relative angular distance between the three poles (the poles of angular momentum, of figure and of rotation) is supposed to be very small, as is the case for the Earth. Throughout this study we will consider that the three poles coincide. So we denote $\overrightarrow{P_{v}}$ as the unit vector directed towards the pole of Venus at $t=0$ defined by :
\begin{eqnarray}
\overrightarrow{P_{v}}=\begin{pmatrix}
\cos l_{p}^{0} \cos b_{p}^{0}\\  \\ \sin l_{p}^{0} \cos b_{p}^{0} \\ \\ \sin b_{p}^{0}
\end{pmatrix} = \mathtt{\textbf{P}} \cdot \begin{pmatrix}
\cos \delta_{p}^{0}  \cos \alpha_{p}^{0} \\ \\ \sin \alpha_{p}^{0} \cos \delta_{p}^{0} \\  \\ \sin \delta_{p}^{0}
\end{pmatrix}
\end{eqnarray}
where $l_{p}^{0}$ and $b_{p}^{0}$ are the ecliptic coordinates, respectively the longitude and the latitude of the pole of Venus. $\mathtt{\textbf{P}}$ is the matrix that converts equatorial coordinates to ecliptic coordinates. $\alpha_{p}^{0},\delta_{p}^{0}$ are the equatorial coordinates of the Venusian north pole with respect to the equator at J2000.0. We take $(\alpha_{p}^{0},\delta_{p}^{0})$ according to the IAU report on cartographic coordinates(1991)(Habibullin, 1995).
\begin{equation}
\alpha_{p}^{0}=272.76 \quad \delta_{p}^{0} = 67.16.
\end{equation}
To determine $\vernal_{0V}$ we use the following equations (see Fig.\ref{OPV}.) :  
\begin{equation}\label{obliquite}
\cos I_{0} = \overrightarrow{P_{ov}} \cdot \overrightarrow{P_{v}} 
\end{equation}
\begin{equation}\label{vern}
\sin I_{0}  \frac{ \overrightarrow{0\vernal_{V}}}{|0\vernal_{V}|} = \overrightarrow{P_{v}} \wedge \overrightarrow{P_{ov}} \Rightarrow \overrightarrow{u}= \frac{\overrightarrow{0\vernal_{V}}}{|0\vernal_{V}|} = \frac{\overrightarrow{P_{v}} \wedge \overrightarrow{P_{ov}}}{\sin I_{0}} 
\end{equation}
where $P_{v}$ and $P_{ov}$ are the unit vectors described previously, $I_{0}$ is the obliquity (angle between the axis of angular momentum and the 0Z inertial axis) and $\overrightarrow{u}$ is the unit vector along $\overrightarrow{0\vernal_{0V}}$.

We finally obtain :
\begin{equation}
I_{0}= 2.634 , \ b_{p}^{0}=-0.103 , \ l_{p}^{0}=57.75. 
\end{equation}

\subsection{Determination of $\Pi_{1}$ and $\pi_{1}$}\label{4.3}

\begin{figure}[htbp]
\center
\resizebox{1.\hsize}{!}{\includegraphics{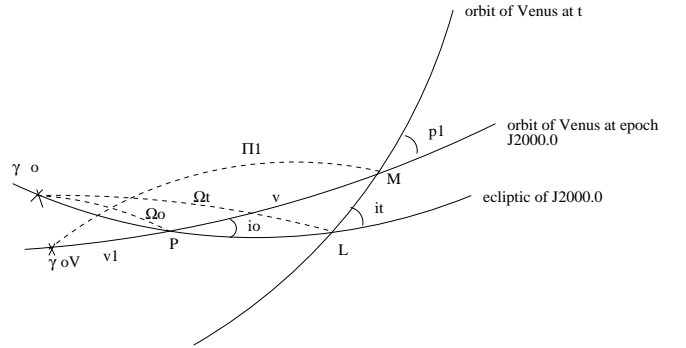}}
 \caption{Motion of the orbit of Venus at the date $t$ with respect to the same orbit at J2000.0.}
 \label{Pi}
\end{figure}

Using the previous section we can now determine the angles $\Pi_{1}$ and $\pi_{1}$, which  are respectively the longitude of the ascending node and the angle between the two orbital planes. The angle $\Pi_{1}$ (see Fig.\ref{PL2}.) is equal to the sum of the angle $v$ and the angle $v_{1}$ (see Fig.\ref{Pi}.) :
\begin{equation}\label{pi1}
\Pi_{1} = v_{1}+ v
\end{equation}

The relationship between $\pi_{1}$, $v$, $i_{0}$, $i_{t}$, $\Omega_{0}$ and $\Omega_{t}$ are derived from the spherical triangle (P,L,M) :

\begin{equation}\label{j}
\cos \pi_{1} =\cos i_{0} \cos i_{t} +\sin i_{0} \sin i_{t} \cos (\Omega_{t}-\Omega_{0})
\end{equation}
\begin{equation}\label{k}
\cos v \sin \pi_{1} = -\sin i_{0} \cos i_{t} +\cos i_{0} \sin i_{t} \cos (\Omega_{t}-\Omega_{0})
\end{equation}
 \begin{equation}\label{l}
\sin v \sin \pi_{1} =\sin i_{t} \sin (\Omega_{t}-\Omega_{0})
\end{equation}
where $ i_{0}$, $i_{t}$ are respectively the inclinations of the orbit of Venus at J2000.0 and at the date $t$ with respect to the ecliptic at J2000.0. $\Omega_{0}$ et $\Omega_{t}$ are the longitude of the ascending node respectively at J2000.0 and at the date $t$. Simon et al. (1994) gave :
\begin{eqnarray*}
i_{t}=3.39466189 -30".84437t -11".6783 t^2+0".03338t^3 \\
\Omega_{t} =76.67992019 - 10008".4815t -51"3261t^{2}-0".58910t^3
\end{eqnarray*}
where $t$ is counted in thousands of Julian years starting from J2000.0.  In this section we choose to carry our calculations to  third order in time.
The relationships between $v_{1}$,$b_{p}^{0}$ and $i_{0}$ are derived from the spherical triangle ($\vernal_{OV}$GP) (see Fig.\ref{AGPi}.):
\begin{figure}[htbp]
\center
\resizebox{0.7\hsize}{!}{\includegraphics{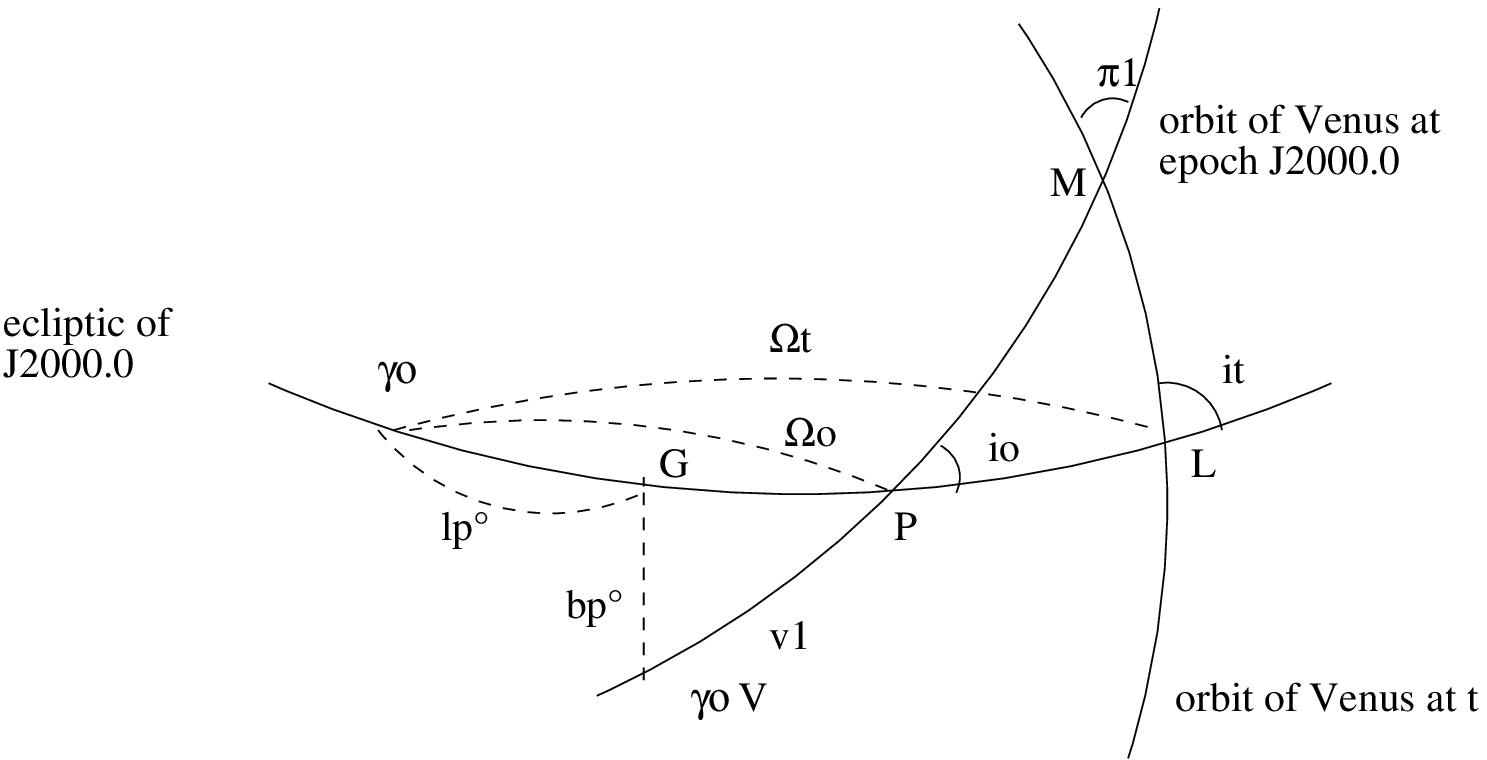}}
 \caption{Motion of the orbit of Venus at the date $t$ with regard to the same orbit at J2000.0.}
\label{AGPi}
\end{figure}
\begin{equation}
\sin v_{1} \sin i_{0}= \sin b_{p}^{0}.
\end{equation}
Using the equations (\ref{j}), (\ref{k}), (\ref{l}) we obtain :
\begin{equation*}
v_{1}=18.92
\end{equation*}
\begin{eqnarray}
\Pi_{1} (t)\nonumber &=&v_{1}+v = 381348".-901."346 \quad t-0.0910822 \quad t^{2}\nonumber \\&&-0".000759571 \quad t^{3}+O(t^4)\\\nonumber
\end{eqnarray}
\begin{equation}
\pi_{1}(t)=59".2626  \quad t +0.0285354  \quad t^{2}+0.00028575  \quad t^{3}+O(t^4)
\end{equation}
where $t$ is counted in thousands of Julian years starting from J2000.0.
For the Earth we have : 
\begin{eqnarray}
\Pi_{1E} (t)
&=&629543." -867.927\quad t+0".1534\quad t^{2}\\ \nonumber&&+0".0000053 \quad t^{3}+O(t^4)\\\nonumber
\end{eqnarray}
\begin{eqnarray}
\pi_{1E}(t)&&=46".9973\quad t -0".0335053\quad t^{2}\nonumber \\&&
-0".00012374\quad t^{3}+O(t^4).
\end{eqnarray}
The relative motion of the orbital plane at $t$ with respect to the orbital plane at $t$=0 is significantly greater for Venus than for the Earth. We used here the mean orbital elements given by Simon et al.(1994). These elements are valid over 3000 years with a relative accuracy of $10^{-5}$ and thus so too are our results

\section{The Hamiltonian of the system and the equation of the motion of rotation}\label{5}

\subsection{The equation of the motion of rotation}
As explained in section \ref{4}, the Hamiltonian related to the rotational motion of Venus is :
\begin{equation}
K"=F_{o}+ E + E'+U
\end{equation}
where $F_{o}$ is the Hamiltonian for the free motion, $E+E'$ is a component related to the motion of the orbit of Venus, which is caused by planetary perturbations. The expressions of $F_{o}$ and $E+E'$ have been set in the previous section. $U$ is the disturbing potential due to external disturbing bodies. Here the external disturbing body is the Sun (The perturbation due to the planets can be neglected), and its disturbing potential is given by :
\begin{equation*}
U = U_{1}+U_{2}
\end{equation*}

\begin{equation}\label{ddd}
U_{1}=\frac{\mathtt{\textbf{G}} M'}{r^3}[\frac{2C-A-B}{2}P_{2}(\sin \delta )+\frac{A-B}{4}P_2^{2} (\sin \delta) \cos 2\alpha]
\end{equation}
\begin{eqnarray}
&U_{2}&=\sum_{n=3}^\infty \frac{\mathtt{\textbf{G}} M'M_{V}a_{V}^n}{r^{n+1}}[J_{n}P_{n}(\sin \delta)\nonumber\\&&-\sum_{m=1}^n{P_{n}^m (\sin\delta)\cdot (C_{nm}\cos m\alpha +S_{nm} \sin m\alpha) }]
\end{eqnarray}
where $\mathtt{\textbf{G}}$ is the gravitationnal constant, $M'$ is the mass of the Sun, $r$ is the distance between its barycenter and the barycenter of Venus. $\alpha$ and $\delta$ are respectively the planetocentric longitude and latitude of the Sun (not to be confused with the usual equatorial coordinates), with respect to the mean equator of Venus and with respect to a meridian origin. The $P_{n}^m$ are the classical Legendre functions given by :
\begin{equation}\label{legendre}
P_{n}^m (x)= \frac{(-1)^m(1-x^2)^{\frac{m}{2}}}{2^n n!}\frac{d^{n+m} (x^2-1)^n}{d^{n+m}x}.
\end{equation}
The Hamiltonian equations are :
\begin{eqnarray}\label{ecano1}
\frac{d}{dt}(L',G',H')= -\frac{\partial K"}{\partial(l',g',h')}
\end{eqnarray}
\begin{eqnarray}\label{ecano2}
\frac{d}{dt}(l',g',h')= \frac{\partial K"}{\partial(L',G',H')}. 
\end{eqnarray}
In the following  the prime notations used above are omitted for the sake of clarity. Using the following equations seen in section \ref{3} :
\begin{equation}
\cos I = \frac{H}{G} \qquad \mathtt{and} \qquad  \cos J= \frac{L}{G}.
\end{equation}
The canonical equations (\ref{ecano1}) and (\ref{ecano2}) become (Kinoshita, 1977):
\begin{eqnarray}\label{e1}
\frac{dl}{dt}= -\frac{1}{G} \sin J \frac{\partial K}{\partial J}
\end{eqnarray}
\begin{eqnarray}\label{e2}
\frac{dg}{dt}= \frac{1}{G}[ \cot J \frac{\partial K}{\partial J}+ \cot I\frac{\partial K}{\partial I}]
\end{eqnarray}
\begin{eqnarray}\label{e3}
\frac{dh}{dt}=-\frac{1}{G\sin I} \frac{\partial K}{\partial I} \label{e3}
\end{eqnarray}
and
\begin{eqnarray}\label{e4}
\frac{dI}{dt}&&= \frac{1}{G}[ \cot I \frac{dG}{dt}-\frac{1}{\sin I} \frac{dH}{dt}]\nonumber\\&&= \frac{1}{G}[\frac{1}{\sin I} \frac{\partial K}{\partial h}-\cot I\frac{\partial K}{\partial g}]. 
\end{eqnarray}
Replacing $I$ by $J$ in the precedent equation yields the variation of $J$.  $I$ is the obliquity and $J$ is the small angle between the angular momentum axis and the figure axis.
To solve the six differential equations (\ref{ecano1}) and (\ref{ecano2}), Kinoshita (1977) used the Hori's method (1966). Here we solve a simplified system, i.e the potential $U_{2}$ does not appear in our study.
To know the motion of precession and nutation of Venus, we have to solve only the following equations :
\begin{eqnarray}
\frac{dI}{dt}&=&\frac{1}{G}[ \cot I \frac{dG}{dt}-\frac{1}{\sin I} \frac{dH}{dt}]\nonumber\\&&= \frac{1}{G}[\frac{1}{\sin I} \frac{\partial K}{\partial h}-\cot I\frac{\partial K}{\partial g}] \\
\frac{dh}{dt}&=& -\frac{1}{G\sin I} \frac{\partial K}{\partial I} 
\end{eqnarray}
where $I$ characterizes the obliquity and $h$ the motion of precession-nutation in longitude.
 These equations can also be written as :
\begin{eqnarray}\label{e6}
\frac{dI}{dt}&&= \frac{1}{G}[ \cot I\frac{dG}{dt}-\frac{1}{\sin I} \frac{dH}{dt}]\nonumber\\&&= \frac{1}{G}[\frac{1}{\sin I} \frac{\partial U_{1}}{\partial h}-\cot I\frac{\partial U_{1}}{\partial g}] 
\end{eqnarray}
\begin{eqnarray}\label{e7}
\frac{dh}{dt}= -\frac{1}{G\sin I}  \frac{\partial U_{1}}{\partial I}
\end{eqnarray}
Integrating these equations, we obtain the variations of the angles $h$ and $I$ :
\begin{eqnarray}\label{wx}
\Delta I =\frac{1}{G}[\frac{1}{\sin I} \frac{\partial}{\partial h} \int U_{1}dt-\cot I\frac{\partial}{\partial g}\int U_{1} dt]
\end{eqnarray}
\begin{eqnarray}\label{wxx}
\Delta h =-\frac{1}{G\sin I}  \frac{\partial}{\partial I}{\int U_{1}dt}.
\end{eqnarray}

\subsection{Development of the disturbing function}

The disturbing potential (\ref{ddd}) is a functions of the modified Legendre polynomials $P_{2}(\sin \delta)$ and $P_2^{2}(\sin \delta)$ and of the planetocentric longitude $\alpha$ and latitude $\delta$ of the Sun, with respect to the mean equator of Venus and with respect to a meridian origin. From the ephemerides we only know the longitude $\lambda$ and the latitude $\beta$ of the Sun with respect to the orbit of Venus. So we use the transformation described by Kinoshita (1977) and based on the Jacobi polynomials, which expresses $\alpha$ and $\delta$ as a function of $\lambda$ and $\beta$. Applying this transformation we obtain:
\begin{eqnarray}\label{aaa}
P_{2}(\sin \delta)&\nonumber& = \frac{1}{2}(3\cos^2 J-1)\Bigg[\frac{1}{2}(3\cos^2 I-1)P_{2}(\sin \beta)\nonumber \\&&-\frac{1}{2} \sin 2I  P_{2}^1(\sin \beta)\cos 2(\lambda - h)\Bigg]\nonumber \\&&+\sin 2J \Bigg[-\frac{3}{4}\sin 2I P_{2}(\sin \beta)\cos g \nonumber \\ &&-\sum_{\epsilon =\pm 1}\frac{1}{4}(1+\epsilon \cos I)(-1+2\epsilon \cos I) \nonumber \\&& P_{2}^1(\sin \beta)\sin(\lambda-h-\epsilon g) \nonumber \\&& -\sum_{\epsilon =\pm 1}\frac{1}{8}\epsilon \sin I(1+\epsilon \cos I)\nonumber \\&& P_{2}^2(\sin \beta)\cos(2\lambda-2h-\epsilon g)\Bigg]+\sin^2 J \nonumber \\&& \Bigg[\frac{3}{4}\sin^2I P_{2}(\sin \beta)\cos 2g+\frac{1}{4}\sum_{\epsilon =\pm 1}\epsilon \sin I  \nonumber \\&& (1+\epsilon \cos I)P_{2}^1(\sin \beta)\sin(\lambda-h-2\epsilon g)-\frac{1}{16}\nonumber \\&& \sum_{\epsilon =\pm 1}(1+\epsilon \cos I)^2P_{2}^2(\sin \beta)\cos 2(\lambda-h-\epsilon g)\Bigg]\nonumber \\
\end{eqnarray}
and
\begin{eqnarray}
&P_{2}^2(\sin \delta)\nonumber&\cos 2\alpha =3\sin^2 J\Bigg[-\frac{1}{2}(3\cos^2 I-1)P_{2}(\sin \beta)\nonumber \\&& \cos 2l+\frac{1}{4}\sum_{\epsilon =\pm 1}\sin 2IP_{2}^1(\sin \beta)\sin(\lambda-h-2\epsilon l) \nonumber \\&& +\frac{1}{8}\sin^2 IP_{2}^2(\sin \beta)\cos 2(\lambda-h-\epsilon l)\Bigg]\nonumber \\ &&+\sum_{\rho =\pm 1}\rho \sin J (1+\rho \cos J)\nonumber \\&& \Bigg[-\frac{3}{2}\sin 2IP_{2}(\sin \beta) \cos (2\rho l+g)\nonumber \\&&-\sum_{\epsilon =\pm 1}\frac{1}{2}(1+\epsilon \cos I)(-1+2\epsilon \cos I)\nonumber \\&&  P_{2}^1(\sin \beta)\sin(\lambda-h-2\rho \epsilon l-\epsilon g)\nonumber \\&& -\sum_{\epsilon =\pm 1}\frac{1}{4}\epsilon \sin I(1+\epsilon \cos I) \nonumber \\&&  P_{2}^2(\sin \beta)\cos(2\lambda-2h-2\rho\epsilon l-\epsilon g)\Bigg]\nonumber
\end{eqnarray}
\begin{eqnarray}\label{aaaa}
&&+\sum_{\rho =\pm 1}\frac{1}{4}(1+\rho \cos J)^2\Bigg[-3\sin^2 IP_{2}(\sin\beta)\cos(2l+2\rho g)\nonumber \\&&-\sum_{\epsilon =\pm 1}\epsilon \sin I(1+\epsilon \cos I)\nonumber \\&& P_{2}^1(\sin\beta) \sin(\lambda -h-2 \rho \epsilon l-2\epsilon g)\nonumber \\&&+\sum_{\epsilon =\pm 1}\frac{1}{4}(1+\epsilon \cos I)^2 \nonumber \\&&  P_{2}^2(\sin\beta)\cos 2(\lambda-h-\rho\epsilon l- \epsilon g)\Bigg].\nonumber \\
\end{eqnarray}
Using (\ref{legendre}) we obtain :
\begin{equation*}
P_{2}(\sin\beta)=\frac{1}{2}(3\sin ^2\beta -1)
\end{equation*}
\begin{equation*}
P_{2}^1(\sin\beta)=-3\sin \beta (1-\sin ^2\beta )^{\frac{1}{2}}
\end{equation*}
\begin{equation*}
P_{2}^2(\sin\beta)=3(1-\sin ^2\beta )
\end{equation*}
where $\beta$ is the latitude of the disturbing body with respect to the orbit of Venus at the date $t$. In our problem the disturbing body is the Sun, so that $\beta =0$. This yields :
\begin{equation*}
P_{2}(\sin\beta)=-\frac{1}{2}
\end{equation*}
\begin{equation*}
P_{2}^1(\sin\beta)=0
\end{equation*}
\begin{equation*}
P_{2}^2(\sin\beta)=3.
\end{equation*}

Assuming that the angle $J$ is small ($J\approx 0$),as it the case for the Earth, (\ref{aaa}) and (\ref{aaaa}) can be written :
\begin{eqnarray}\label{relation3}
P_{2}(\sin \delta) = [-\frac{1}{4}(3\cos ^2 I-1)-\frac{3}{4}\sin^2 I \cos 2 (\lambda-h)]
\end{eqnarray}
and
\begin{eqnarray}\label{relation4}
P_{2}^2(\sin \delta)\cos 2\alpha=[ \frac{3}{2} \sin^2I \cos(2l+2g)\nonumber \\ +\sum_{\epsilon=\pm 1}\frac{3}{4}(1+\epsilon \cos I)^2\cdot \cos 2(\lambda-h-\epsilon l-\epsilon g)].
\end{eqnarray}
Thus the disturbing potential is simplified in the following form :
\begin{eqnarray}\label{potentiel2}
U_{1}\nonumber&=&\frac{\mathtt{\textbf{G}} M'}{r^3}\Bigg[\frac{2C-A-B}{2}\Big(-\frac{1}{4}(3\cos ^2 I-1)\nonumber\\&&-\frac{3}{4}\sin^2 I \cos 2 (\lambda-h)\Big)\nonumber \\&& +\frac{A-B}{4}\bigg[ \frac{3}{2} \sin^2I \cos(2l+2g)\nonumber \\&& +\sum_{\epsilon=\pm 1}\frac{3}{4}(1+\epsilon \cos I)^2\cdot \cos 2(\lambda-h-\epsilon l-\epsilon g)\bigg]\Bigg].\nonumber \\
\end{eqnarray}

\section{Determination of the precession and the nutation of Venus\label{6}}

Replacing $U_{1}$ in (\ref{wx}) and (\ref{wxx}) we obtain :
\begin{eqnarray}\label{e8}
&\Delta h\nonumber & =-\frac{1}{G\sin I} \frac{\partial}{\partial I} \int \bigg(\frac{\mathtt{\textbf{G}} M'}{r^3}[\frac{2C-A-B}{2}\nonumber \\
&&\Big[-\frac{1}{4}(3\cos^2 I-1)-\frac{3}{4}\sin^2 I \cos 2 (\lambda -h)\Big]\nonumber \\
&&+\frac{A-B}{4}\Big[ \frac{3}{2} \sin^2l \cos(2l+2g)\nonumber \\
&& +\sum_{\epsilon=\pm 1}\frac{3}{4}(1+\epsilon \cos I)^2\nonumber \\
&&  \cos 2(\lambda -h-\epsilon l-\epsilon g)\Big]\bigg]\Bigg)dt\nonumber \\
\end{eqnarray}
\begin{eqnarray}\label{e9}
\Delta I\nonumber& =&\frac{1}{G}\Bigg[\frac{1}{\sin I} \frac{\partial}{\partial h} \int \Bigg(\frac{\mathtt{\textbf{G}} M'}{r^3}\bigg[\frac{2C-A-B}{2}\nonumber \\&& \Big[-\frac{1}{4}(3\cos ^2 I-1)-\frac{3}{4}\sin^2 I \cos 2 (\lambda-h)\Big]\nonumber \\&&+\frac{A-B}{4}\Big[\frac{3}{2} \sin^2I \cos(2l+2g)\nonumber \\&& +\sum_{\epsilon=\pm 1}\frac{3}{4}(1+\epsilon \cos I)^2\nonumber \\&& \cos 2(\lambda-h-\epsilon l-\epsilon g)\Big]\bigg]\Bigg)dt\nonumber \\&&
-\cot I \frac{\partial}{\partial g}\int \Bigg(\frac{\mathtt{\textbf{G}} M'}{r^3}\Bigg[\frac{2C-A-B}{2} \nonumber \\&&  \Big[-\frac{1}{4}(3\cos ^2 I-1)-\frac{3}{4}\sin^2 I \cos 2 (\lambda-h)\Big]\nonumber \\&&+\frac{A-B}{4}\Big[\frac{3}{2} \sin^2I \cos(2l+2g)\nonumber \\ &&+\sum_{\epsilon=\pm 1}\frac{3}{4}(1+\epsilon \cos I)^2\nonumber \\&&  \cos 2(\lambda-h-\epsilon l-\epsilon g)\Big]\Bigg]\Bigg)dt\Bigg].\nonumber \\
\end{eqnarray}
To simplify the calculations, we study separately the components depending on the dynamical flattening and those depending on the triaxiality of Venus. As the dynamical flattening corresponds to the symmetric part at the right hand side of (\ref{e8}) and (\ref{e9}) although the triaxiality corresponds to the antisymmetric one, in the following the coefficients depending on dynamical flattening will be denoted with an index "s" and the coefficients depending on the triaxility with an index "a". Moreover the expression $\frac{\mathtt{\textbf{G}}M'}{r^3}$ in (\ref{e8}) and (\ref{e9}) can be replaced by $\frac{\mathtt{\textbf{G}}M'}{a^3} \cdot \frac{a^3}{r^3}$ where a is the semi-major axis of Venus given by :
\begin{equation}
n^2a^3 = \mathtt{\textbf{G}} M'.
\end{equation}
 Finally we have the following equations :
\begin{eqnarray}\label{osc1}
\Delta hs =-\frac{1}{G\sin I} \frac{\partial}{\partial I} \int W_{1} dt
\end{eqnarray}
\begin{eqnarray}\label{osc2}
\Delta Is& =&\frac{1}{G}\frac{1}{\sin I} \frac{\partial}{\partial h} \int W_{1} dt 
\end{eqnarray}
and :
\begin{eqnarray}\label{osc3}
\Delta ha& =&-\frac{1}{G\sin I}  \frac{\partial}{\partial I} \int W_{2} dt
\end{eqnarray}
\begin{eqnarray}\label{osc4}
\Delta Ia &=&\frac{1}{G}\Bigg[\frac{1}{\sin I} \frac{\partial}{\partial h} \int W_{2} dt -\cot I \frac{\partial}{\partial g} \int W_{2} dt \Bigg]
\end{eqnarray}
where
\begin{eqnarray}
&W_{1}&=\frac{\mathtt{\textbf{G}}M'}{a^3}(\frac{a}{r})^3\Big(\frac{2C-A-B}{2}\big[-\frac{1}{4}(3\cos ^2 I-1)\nonumber\\&&-\frac{3}{4}\sin^2 I \cos 2 (\lambda-h)\big]\Big)
\end{eqnarray}
\begin{eqnarray}
W_{2}&=&\frac{\mathtt{\textbf{G}}M'}{a^3}(\frac{a}{r})^3\bigg(\frac{A-B}{4}\Big[\frac{3}{2} \sin^2I \cos(2l+2g)\nonumber \\ &&+\sum_{\epsilon=\pm 1}\frac{3}{4}(1+\epsilon \cos I)^2 \nonumber \\&& \cos 2(\lambda-h-\epsilon l-\epsilon g)\Big]\bigg).
\end{eqnarray}
We adopt the following notations :
\begin{equation}\label{dodo}
\Delta \psi= \Delta h = (\Delta hs+\Delta ha)
\end{equation}
\begin{equation}\label{titi}
\Delta \epsilon = \Delta I = (\Delta Is+\Delta Ia).
\end{equation}
As the rotation of Venus is retrograd the convention used is not the same as IAU to the Earth.

\section{Development of the disturbing function}\label{7}
Considering the level of accuracy and the time interval involved (3000 y), truncating the function $e$ to first order in time is a sufficient approximation. Therefore we take the eccentricity of Venus from Simon et al.(1994) as :
\begin{equation}
e=0.0067719164-0.0004776521t.
\end{equation}

\subsection{Terms depending on the dynamical flattening}\label{lo}

To solve the equations (\ref {osc1}) and (\ref{osc2}) which give the nutation coming from the dynamical flattening, it is necessary to develop $\frac{1}{2}(\frac{a}{r})^3$ and $(\frac{a}{r})^3 \cos 2 (\lambda-h)$ with respect to time (i.e with respect to the mean anomaly and the mean longitude of Sun). Using Kepler's equation :
\begin{equation}\label{kepler}
u-e\sin u = n (t-t_{0})=M
\end{equation}
where $u$ is the eccentric anomaly, $e$ the eccentricity, $n$ the mean motion and $M$ the mean anomaly of Venus, we have the following classical development:
\begin{eqnarray}\label{ab}
\frac{1}{2}(\frac{a}{r})^3&&=\frac{1}{2} (1+\frac{3}{2}e^2)+\frac{1}{2}(3e+\frac{27}{8}e^3)\cos M \nonumber \\&&+\frac{9}{4}e^2 \cos 2M+\frac{53}{8}e^3 \cos 3M.
\end{eqnarray}
For comparison we also give the corresponding coefficients of the development of the Earth (see Table \ref{dev1}).
\begin{table}[!h]
\begin{center}
\resizebox{1.\hsize}{!}{\begin{tabular}[h]{lcccc}

 M  & $L_{S}$ & period &&   \\
 & &days &  $\cos \times 10^{-7}$ & $t \cos \times  10^{-7}$  \\ 
\hline\\

0& 0 & &$\frac{1}{2}(1+\frac{3}{2}e^2)=5000344~~[5002093]$ &-96 ~~[-105] \\
\\
 1 & 0 & 224,70~~[365,26] & $\frac{1}{2}(3e+\frac{27}{8}e^3)= 101584~~[250708]$  & -7164~~[-6305]\\
\\
 2 & 0 & 112,350~~[182,63] & $\frac{9}{4}e^2=1032~~[6282]$ & -146~~[-316] \\
\\
3&0&74.900~~[121.753]& $\frac{53}{8}e^3=200~~[308]$&-40~~ [-23]\\
\hline
\end{tabular}
}
\end{center}
\caption{Development of $\frac{1}{2} \left( \frac{a}{r} \right) ^3 $ of Venus (the corresponding values for the Earth are in square brackets). $t$ is counted in Julian centuries.}
\label{dev1}
\end{table}
 We have the following definition :
\begin{equation}
2\lambda = 2(\overline{\omega} + v) =2(\overline{\omega}+M+(v-M))=2L_{S}+2(v-M)
\end{equation}
where $\lambda$ is the true longitude of the Sun, $\overline{\omega}$ the longitude of the periapse of Venus, $M$ the mean anomaly and $L_{S}$ the mean longitude of the Sun.
From Kepler's equation we have the following classical development :
\begin{equation}\label{vm}
v-M = (2e-\frac{e^3}{4}) \sin M +(\frac{5}{4}e^2-\frac{11}{24}e^4)\sin 2M + \frac{13}{12}e^3\sin 3M.
\end{equation}
From the trigonometric equation :
\begin{equation}
\cos 2\lambda =\cos 2 L_{S} \cos 2(v-M)-\sin 2L_{S} \sin 2(v-M)
\end{equation}
we obtain :
\begin{eqnarray}\label{dla}
(\frac{a}{r})^3 \cos 2 (\lambda-h)&&=(\frac{a}{r})^3(\cos 2 L_{S} \cos 2(v-M)\nonumber\\&&-\sin 2L_{S} \sin 2(v-M)).
\end{eqnarray}
Using the developments (\ref{vm}) and (\ref{dla}) and the classical trigonometric relationships we obtain a development of $(\frac{a}{r})^3 \cos 2 (\lambda-h)$ :
\begin{eqnarray}\label{cd}
(\frac{a}{r})^3 \cos 2 (\lambda-h)\nonumber &&=(1-\frac{5}{2}e^2)\cos 2 L_{S}\nonumber\\&&+(\frac{7}{2}e-\frac{123}{16}e^3)\cos (2 L_{S} +M)\nonumber \\&& +(-\frac{1}{2}e+\frac{1}{16}e^3)\cos (2 L_{S} -M)\nonumber\\&&+(\frac{17}{2}e^2)\cos (2 L_{S}+2M)\nonumber\\&&+\frac{845}{48}e^3 \cos (2L_{S}+3M)\nonumber\\&&+\frac{1}{48}e^3 \cos(2L_{S}-3M).\nonumber \\
\end{eqnarray}
The coefficients of this development are given in Table \ref{dev2}.
\begin{table}[!h]
\begin{center}
\resizebox{1.\hsize}{!}{\begin{tabular}[h]{lcccc}

    M & $L_{S}$ & period & &   \\
    & &days & $\cos \times 10^{-7}$ & $t\cos \times 10^{-7}$  \\
\hline \\
 0 & 2 & 112,35 ~~[182,62] &$(1-\frac{5}{2}e^2)= 9998853~~[9993025] $ & 161~~ [351]  \\
\\
-1 & 2 & 224,70~~[365,22] & $(-\frac{1}{2}e+\frac{1}{16}e^3)=-33859~~[-83540]$ & 2388~~[2101] \\
\\
1 & 2 & 74,90~~[121,75]  &$(\frac{7}{2}e-\frac{123}{16}e^3)= 236993~~[584444] $& -16718~~[- 14713] \\
\\
 2 & 2 & 56,17~~[91,31] & $(\frac{17}{2}e^2)=3898~~[23730]$ & -550~~[-1193]\\
\\
3&2&44.94~~[73.05]&$\frac{845}{48}e^3=54~~[821]$&-4~~[-20]\\
\\
-3&2&-224.70~~[-365.25]&$\frac{1}{48}e^3=0[0]$&0[0]\\

\hline
\end{tabular}
}
\end{center}
\caption{Development of $ \left( \frac{a}{r} \right) ^3 \ \cos(2( \lambda-h)$ of Venus (the corresponding values for the Earth are in square brackets). $t$ is counted in Julian centuries.}
\label{dev2}
\end{table}
The periods in Tables \ref{dev1} and \ref{dev2} have been computed through the mean elements of the planets given by the ephemerides VSOP87 of Simon et al.(1994). Because the eccentricity of Venus is smaller than the eccentricity of the Earth, the coefficients of the development for Venus are smaller than the corresponding ones for the Earth, except for the leading one in Table \ref{dev2}.

\subsection{Terms depending on the triaxiality}\label{tata}

To solve the equations (\ref {osc3}) and (\ref{osc4}) it is necessary to develop $\frac{1}{2}(\frac{a}{r})^3\cos 2\Phi$, ($\frac{a}{r})^3\cos (2(\lambda-h)  - 2\Phi)$ and $(\frac{a}{r})^3\cos (2(\lambda-h) + 2\Phi)$ with respect to the mean anomaly $M$, the mean longitude $L_{S}$ of Sun and to the angle $l+g\approx \Phi$ according to  (\ref{didi}). Using classical trigonometric equations and developments similar to those used in section \ref{lo}, we obtain the following :
\begin{eqnarray}\label{ef}
(\frac{a}{r})^3 \cos 2 (\lambda-h-\Phi)\nonumber &&=(1-\frac{5}{2}e^2)\cos (2 L_{S}-2\Phi)\nonumber\\&&+(\frac{7}{2}e-\frac{123}{16}e^3)\cos (2 L_{S} +M-2\Phi)\nonumber \\&& +(-\frac{1}{2}e+\frac{1}{16}e^3)\cos (2 L_{S} -M-2\Phi)\nonumber\\&&+(\frac{17}{2}e^2)\cos (2 L_{S}+2M-2\Phi)\nonumber \\ &&+\frac{845}{48}e^3 \cos (2L_{S}+3M-2\Phi)\nonumber\\&&+\frac{1}{48}e^3 \cos(2L_{S}-3M-2\Phi)\nonumber \\
\end{eqnarray}
\begin{eqnarray}\label{gh}
(\frac{a}{r})^3 \cos 2 (\lambda-h+\Phi)\nonumber&&=(1-\frac{5}{2}e^2)\cos (2 L_{S}+2\Phi)\nonumber\\&&+(\frac{7}{2}e-\frac{123}{16}e^3)\cos (2 L_{S} +M+2\Phi)\nonumber \\&& +(-\frac{1}{2}e+\frac{1}{16}e^3)\cos (2 L_{S} -M+2\Phi)\nonumber\\&&+(\frac{17}{2}e^2)\cos (2 L_{S}+2M+2\Phi)\nonumber \\&&+\frac{845}{48}e^3 \cos (2L_{S}+3M+2\Phi)\nonumber\\&&+\frac{1}{48}e^3 \cos(2L_{S}-3M+2\Phi)\nonumber \\
\end{eqnarray}
\begin{eqnarray}\label{ij}
\frac{1}{2}(\frac{a}{r})^3 \cos (2\Phi)\nonumber&&=(\frac{1}{2}+\frac{3}{4}e^2)\cos (2\Phi)\nonumber\\&&+(\frac{3}{4}e+\frac{27}{32}e^3)\cos(M-2\Phi)\nonumber \\ && +(\frac{3}{4}e+\frac{27}{32}e^3)\cos (M+2\Phi)\nonumber\\&&+\frac{9}{8}e^2\cos (2M-2\Phi)\nonumber \\ && +(\frac{9}{8}e^2)\cos (2M+2\Phi)\nonumber \\&&+\frac{53}{16}e^3 \cos (3M+2\Phi)\nonumber\\&&+\frac{53}{16}e^3 \cos(3M-2\Phi).\nonumber \\
\end{eqnarray}
These are given in Tables \ref{dev3},\ref{dev4} and \ref{dev5}
\begin{table}[!h]
\begin{center}
\resizebox{.8\hsize}{!}{\begin{tabular}[h]{lccccc}

    M & $L_{S}$ &$\Phi$ & period & &  \\
    & & &days & $\cos \times 10^{-7}$ & $t\cos \times 10^{-7}$   \\
\hline\\

 0 & 0 & 2  & -121.51    &$(\frac{1}{2}+\frac{3}{4}e^2)= 5000344$& -48    \\
\\
 1 & 0 & 2  & -264.6   & $(\frac{3}{4}e+\frac{27}{32}e^3)=50792$  & -3582 \\
\\
 1 & 0 & -2 & 78.86   & $(\frac{3}{4}e+\frac{27}{32}e^3)=50792$  & -3582 \\
\\
 2 & 0 &  2 & 1490.35   & $(\frac{9}{8}e^2)=516$    & -72  \\
\\
 2 & 0 & - 2 & 58.37   & $(\frac{9}{8}e^2)=516$    & -72  \\
\\
3&0&2&195.26&$\frac{53}{16}e^3=10$&0\\
\\
3&0&-2&46.34&$\frac{53}{16}e^3=10$&0\\
\hline
\end{tabular}
}
\end{center}
\caption{Development of $ \left( \frac{a}{r} \right) ^3  \cos(2 \Phi)$ of Venus. $t$ is counted in Julian centuries.}
\label{dev3}
\end{table}
\newpage
\begin{table}[!hbtp]
\begin{center}
\resizebox{.8\hsize}{!}{\begin{tabular}[h]{lcccccc}
  M & $L_{S}$ & $\Phi$ & period &&\\
 &  &  &days & $\cos \times 10^{-7}$ & $t \cos \times  10^{-7}$   \\
\hline \\
 0 & 2 & - 2  &58.37   & $(1-\frac{5}{2}e^2)=9998853$  & 161   \\
\\
 -1 & 2& - 2  &78.87   & $(-\frac{1}{2}e+\frac{1}{16}e^3)=-33859$ & 2388 \\
\\
1 & 2 & - 2  & 46.34  &$(\frac{7}{2}e-\frac{123}{16}e^3)= 236993$ & -16718 \\
\\
 2 & 2 & - 2  & 38.41  & $(\frac{17}{2}e^2)=3898$ & -550\\
\\
-3&2&-2&264.59&$\frac{1}{48}e^3=0$&0\\
\\
3&2&-2&32.80&$\frac{845}{48}e^3=54$&-4\\
\hline
\end{tabular}
}
\end{center}
\caption{Development of $ \left( \frac{a}{r} \right) ^3 \ \cos(2(\lambda-h) - 2 \Phi)$ of Venus. $t$ is counted in Julian centuries.}
\label{dev4}
\end{table}
\newpage
\begin{table}[!hbtp]
\begin{center}
\resizebox{.8\hsize}{!}{\begin{tabular}[h]{lcccccc}
 M & $L_{S}$ & $\Phi$ & period  &&\\
   &  &  &days &$\cos \times 10^{-7}$ & $t\cos \times 10^{-7} $ \\
\hline \\
 0 & 2 &  2  & 1490.35 &$(1-\frac{5}{2}e_{V}^2)=9998853$ & 161   \\
\\
 -1 & 2&  2  & -264.6  & $(-\frac{1}{2}e+\frac{1}{16}e^3)=-33859$ & 2388 \\
\\
 1 & 2 &  2  & 195.26 &$(\frac{7}{2}e-\frac{123}{16}e^3)= 236993$ & -16718 \\
\\
 2 & 2 &  2  & 104.47  &  $(\frac{17}{2}e^2)=3898$ & 550\\
\\
-3&2&2&224.70&$\frac{1}{48}e^3=0$&0\\
\\
3&2&2&71.31&$\frac{845}{48}e^3=54$&-4\\
\hline
\end{tabular}
}
\end{center}
\caption{Development of $ \left( \frac{a}{r} \right) ^3 \ \cos(2(\lambda-h) + 2 \Phi)$ of Venus. $t$ is counted in Julian centuries.}
\label{dev5}
\end{table}
We note that the coefficients in Tables \ref{dev4} and \ref{dev5} are the same as the coefficients of Table \ref{dev2}, because the calculations are similar and they do not depend on the frequency associated with each argument, but only on the eccentricity of Venus. Nevertheless the corresponding periods are different because their calculation includes the argument $\Phi$, the sideral rotation of Venus given by :
\begin{equation}
\Phi\approx l+g=-(\frac{2\Pi}{T_{V}})\qquad t
\end{equation}
where $T_{V}$ is the sidereal period of the rotation of Venus. The numerical value for $T_{V}$ is $T_{V}=243.02$ days.
Notice also that because the rotation is retrograde, $\Phi$ has a negative sign, and that the period of $L_{S}$ (225 d) is close to the period of $\Phi$ (243 d). Therefore the period for the leading componant $2L_{S}+2\Phi$ in Table \ref{dev5} (1490.35 d) is much longer than the period of the leading component $2L_{S}-2\Phi$ in Table \ref{dev4} (58.37 d). We have $L_{S}=\omega_{o} \ t$, $M=\omega_{o\omega} \ t$ and $\Phi=\omega_{r} \ t$ where $\omega_{o}=(2\Pi/224.70080) rd/d$, $\omega_{o\omega}=(2\Pi/224.70082) rd/d$ and $\omega_{r}=-(2\Pi/243.02)  rd/d$. The very small difference between $\omega_{o}$ and $\omega_{o\omega}$ is due to the longitude of periapse $\overline{\omega}$ which corresponds to a very low frequency. As a consequence we do not change our calculations significantly by considering that $\dot{L_{S}}=\dot{M}$ ($ L_{S}=M+ \overline{\omega}$).

\section{Precession and nutation which depend on the longitude and obliquity}\label{8}
 In the following, we calculate the precession and the nutation of Venus depending on  both the dynamical flattening and the dynamical triaxiality.

\subsection{Precession and the nutation of Venus depending on the dynamical flattening}

According to (\ref{osc1}) and (\ref{osc2}) we compute the precession-nutation in longitude ($\Delta h_{s}$) and obliquity ($\Delta I_{s}$) due to the dynamical flattening : 

\begin{eqnarray}\label{bobo}
&\Delta hs& =-\frac{1}{\sin I} \frac{\partial}{\partial I} \int 3 \frac{\mathtt{\textbf{G}}M'}{a^3}\frac{1}{G}\Big(\frac{2C-A-B}{2}\Big)\overline{W_{1}}dt \\\nonumber &&=-\frac{K_{s}}{\sin I} \frac{\partial}{\partial I} \int \overline{W_{1}}dt
\end{eqnarray}
\begin{eqnarray}\label{baba}
&\Delta Is& =\frac{1}{\sin I} \frac{\partial}{\partial h} \int 3 \frac{\mathtt{\textbf{G}}M'}{a^3}\frac{1}{G}\Big(\frac{2C-A-B}{2}\Big)\overline{W_{1}} dt \\\nonumber &&=\frac{K_{s}}{\sin I} \frac{\partial}{\partial h} \int \overline{W_{1}} dt 
\end{eqnarray}
where
\begin{eqnarray}
\overline{W_{1}}=(\frac{a}{r})^3\big[-\frac{1}{12}(3\cos ^2 I-1)-\frac{1}{4}\sin^2 I \cos 2 (\lambda-h)\big].
\end{eqnarray}

$\Delta h_{s}$ and $\Delta I_{s}$ in (\ref{bobo}) and (\ref{baba}) depend directly on the scaling factor $K_{s}$ :

\begin{equation}
K_{s}=3\frac{\mathtt{\textbf{G}}M'}{G a^3}\Bigl[\frac{2C-A-B}{2}\Bigr].
\end{equation}
 Using the third Kepler 's law we obtain :
\begin{equation}
K_{s}=3\frac{n^2}{\omega}\bigl[\frac{2C-A-B}{2C}\bigr]=\frac{3n^2}{\omega} H_{V}
\end{equation}
where $n$ is the mean motion of Venus. The dynamical flattening, defined as  $H_{V}=\frac{2C-A-B}{2C}$ , is a dimensionless parameter. We supposed here to the first order of approximation that the components $\omega_{1}$ and $\omega_{2}$ of the rotation of Venus are negligible with respect to the component $\omega_{3}$ along the figure axis (0,z), as is the case for the Earth (Kinoshita, 1977). So we have 
\begin{equation}
G=A\omega_{1}+B\omega{2}+C\omega_{3}\approx C\omega
\end{equation}
with the values of $A$,$B$,$C$ adopted in section \ref{10} and that of the dynamical flattening given which is deduced directly from them (given in Table \ref{table1}), we find :
\begin{equation}
K_{s} = -8957".55 \pm 133.29 / Julian .\ cy.
\end{equation}
In the case of the Earth, the scaling factor due to the gravitational influence of the Sun is (Souchay et al.1999) :
\begin{equation}
K_{s}^E= 3475".36 / Julian .\ cy.
\end{equation}
Then we have the following result :
\begin{equation}\label{rap}
\frac{K_{s}}{K_{s}^E}=(\frac{n^2}{n_{E}^2})\cdot(\frac{\omega_{E}}{\omega})\cdot (\frac{H_{V}}{H_{E}})\approx -2.577.  
\end{equation}
Thus for Venus, the value of the constant of the precession-nutation due to the Sun is roughly 2.6 times bigger that the corresponding one for the Earth. Through the development done previously (cf section \ref{7}) and according to the conventional notations
 (\ref{dodo}) and (\ref{titi}), we get the precession and the nutation of Venus depending on its dynamical flattening. We will see in the following sub-section that the triaxiality does not contribute to the precession. So we obtain the following result for the precession :
\begin{equation}\label{ppp}
\dot\psi=K_{s}\cos I \int (1+\frac{3}{2}e^2) dt =4474".35t-0.021 t^2
\end{equation}
where $t$ is counted in  Julian centuries . The  nutation coefficients  in longitude and in obliquity are given respectively in Tables \ref{ww1} and \ref{ww2}. All our results are given with respect to the moving orbit at the date $t$.
\begin{table}[!htbp]
\caption{$\Delta \Psi_{s}=\Delta h_{s} $: nutation coefficients in longitude of Venus depending on its dynamical flattening.}
\label{ww1}
\begin{center}
\resizebox{.9\hsize}{!}{\begin{tabular}[htbp]{ccccc}
\hline\\
 Argument & Period & $\sin$ &$t\sin$ & $\cos$ \\
  & days & arc second &arc second/Julian cy &arc second \\
&& $(10^-7)$& $(10^-7)$& $(10^-7)$\\
\hline
$2L_{s}$ & 112.35& 21900468  & 352 & 0 \\
\hline \\
$M$ &224.70 & -889997 &62765 & 61 \\
\hline \\
$ 2L_{s}+M$ & 74.90& 346057&- 24412 &-7\\
\hline\\
$2L_{s}-M$ & 224.70& -148323& 10461 &10\\
\hline \\
$ 2L_{s}+M$ & 56.17& 4269 &- 602& 0 \\
\hline \\
$2M$ & 112.35& -4521 & 640 & 0 \\

\end{tabular}
}
\end{center}
\end{table}

\begin{table}[!htbp]
\caption{$\Delta \epsilon_{s}=\Delta I_{s} $ : nutation coefficients in obliquity of Venus depending on its dynamical flattening.}
\label{ww2}
\begin{center}
\resizebox{.9\hsize}{!}{\begin{tabular}[htbp]{ccccc}
\hline\\
 Argument & Period & $\cos$ &$t\cos$ & $\sin$ \\
  & days & arc second &arc second/Julian century &arc second \\
&& $(10^-7)$& $(10^-7)$& $(10^-7)$\\
\hline\\
$2L_{s }$ & 112.35 & -100741& -16 & 0 \\
\hline \\
$2L_{s}+M$ & 74.90& -15919& 1123& 0 \\
\hline\\
$2L_{s}-M$ & 224.70& 6822& -481& 0\\
\hline \\
$2L_{s}+2M$ & 56.17&-196&27 & 0 \\

\end{tabular}
}
\end{center}
\end{table}
These coefficients are of the same order of magnitude as those for the Earth due to the action of the Sun found by Kinoshita (1977) and Souchay et al. (1999). The largest nutation coefficient has roughly a 2" amplitude whereas for the Earth it has a 1" amplitude. As in the case of the precession, this difference comes from the larger value of the constant $K_{s}$ and is slightly compensated for after integration by the fact that the period of revolution of Venus around the Sun (225 d) is shorter than that of the Earth. The largest nutation coefficient in obliquity is comparatively very small with a 0.1" amplitude whereas for the Earth it has a 0.5" amplitude. The difference is due to the small obliquity of Venus ($I \approx 3$), our results depending on $\sin I$.

\subsection{Precession and nutation of Venus depending on dynamical triaxiality}

Starting from (\ref{osc3}) and (\ref{osc4}), we can compute the nutations $\Delta_{ha}$ and $\Delta{Ia}$ respectively in longitude and in obliquity, due to the triaxiality of Venus : 

\begin{eqnarray}\label{riri}
&\Delta ha & =-\frac{1}{\sin I}  \frac{\partial}{\partial I} \int 3 \frac{\mathtt{\textbf{G}}M'}{a^3}\frac{1}{G}\Big(\frac{A-B}{4}\Big) \overline {W_{2}} dt\\ \nonumber &&= -\frac{K_{a}}{\sin I}  \frac{\partial}{\partial I} \overline {W_{2}} dt
\end{eqnarray}
\begin{eqnarray}\label{roro}
&\Delta Ia\nonumber&=\Bigg[\frac{1}{\sin I} \frac{\partial}{\partial h} \int  3 \frac{\mathtt{\textbf{G}}M'}{a^3}\frac{1}{G}\Big(\frac{A-B}{4}\Big) \overline {W_{2}}dt\nonumber\\&& -\cot I \frac{\partial}{\partial g} \int  3 \frac{\mathtt{\textbf{G}}M'}{a^3}\frac{1}{G}\frac{A-B}{4} \overline {W_{2}} dt \Bigg]\\  && =\frac{K_{a}}{\sin I} \frac{\partial}{\partial h} \int \overline {W_{2}}dt -K_{a} \cot I  \frac{\partial}{\partial g} \int \overline {W_{2}} dt
\end{eqnarray}
where
\begin{eqnarray}
\overline{W_{2}}&=&(\frac{a}{r})^3\ \big[\frac{1}{2} \sin^2I\cdot \cos(2 \Phi)\nonumber \\ &&+\sum_{\epsilon=\pm 1}\frac{1}{4}(1+\epsilon \cos I)^2\nonumber\\&&\cdot \cos 2(\lambda-h-\epsilon \Phi )\big].
\end{eqnarray}

$\Delta ha$ and $\Delta Ia$ in (\ref{riri}) and (\ref{roro}) depend directly on the scaling factor $K_{a}$ : 
\begin{equation}
K_{a}=3\frac{\mathtt{\textbf{G}}M'}{G a^3}\Bigl[\frac{A-B}{4}\Bigr]
\end{equation}
where $M'$, $\mathtt{\textbf{G}}$, $G$ and $a$ have been defined previously. Still using Kepler's third law we get :
\begin{equation}
K_{a}=3\frac{n^2}{\omega}\bigl[\frac{A-B}{4C}\bigr].
\end{equation}
The triaxiality, defined as  $T_{V}=\frac{A-B}{4C}$, is also, as $H_{V}$, a dimensionless parameter. here we assume again that :
\begin{equation}
G\approx C\omega.
\end{equation}
Then, with the value of the triaxiality (given in Table \ref{fig1}), we get :
\begin{equation}
K_{a} = 1131".23 \pm 17.09 \  Julian / cy.
\end{equation}
In the case of the Earth, the scaling factor due to the gravitational influence of the Sun is (Souchay et al.1999) :
\begin{equation}
K_{a}^E= -5"68 \  Julian / cy.
\end{equation}
Thus we have the following ratios :
\begin{equation}\label{rap}
\frac{K_{a}}{K_{a}^E}=(\frac{n_{v}^2}{n_{T}^2})\cdot(\frac{\omega_{E}}{\omega})\cdot (\frac{T_{V}}{T_{E}})\approx -199.16. 
\end{equation}
This result shows that the coefficients of nutation due to the triaxility are considerably larger for Venus than for the Earth. 
Thanks to the development done previously (cf section \ref{tutu}) and according to the equations (\ref{riri}) and (\ref{roro}), we get the nutation of Venus depending on its triaxiality. The  nutation coefficients  in longitude and in obliquity are given respectively in Tables \ref{ww3} and \ref{ww5}. All our results are given with respect to the moving orbit at the date $t$.
\begin{table}[!h]
\caption{$\Delta \Psi_{a}=\Delta h_{a} $: nutation coefficient in longitude of Venus depending on its triaxiality.}
\label{ww3}
\begin{center}
\resizebox{.9\hsize}{!}{\begin{tabular}[htbp]{ccccc}
\hline\\
 Argument & Period & $\sin$ &$t\sin$ & $\cos$ \\
  & days & arc second &arc second/Julian cy &arc second \\
&&$10^-7$&$10^-7$&$10^-7$\\
\hline \\
$2\Phi$ & -121.51 & -5994459 & 58 & 0 \\
\hline\\
$2L_{S}-2\Phi$ &58.37&-2880826& -46& 0 \\
\hline\\
$M+2\Phi$ & -264.6& -132590 & 9351&- 11\\
\hline\\
$M-2\Phi+2L_{S}$&46.34& -54201&3823&0 \\
\hline\\
 $M-2\Phi$ & 78.86& 39519&- 2787& 0 \\
\hline\\
$2L_{S}+2\Phi$ &1490.35& 38866 & 0&0 \\
\hline\\
$-M-2\Phi+2L_{S}$&78.86& 13179&-929&0\\
\hline\\
$2M+2\Phi$ & 1490.35& 7587 &- 1058&-7 \\
\hline\\
$2M-2\Phi+2L_{S}$ &38.41&-739&104& 0\\
\hline\\
$2M-2\Phi$&58.37&297 &-41&0 \\
\hline\\
$M +2\Phi+2L_{S}$ & 195.26& 121&- 8&0 \\
\hline\\
$-M+2\Phi+2L_{S}$&-264.66&23&-2 &0 \\
\hline\\
$2M+2\Phi+2L_{S}$ &104.47&1&0& 0 \\

\end{tabular}

}
\end{center}
\end{table}

\begin{table}[!h]
\caption{$\Delta \epsilon_{a}=\Delta I_{a} $ :nutation coefficients in obliquity of Venus depending on its triaxiality.}
\label{ww5}
\begin{center}
\resizebox{0.9\hsize}{!}{\begin{tabular}[htbp]{ccccc}
\hline\\
 Argument & Period & $\cos$ &$t\cos$ & $\sin$ \\
  & days & arc second &arc second/Julian cy &arc second \\
\hline\\
$2\Phi$ & -121.51 & 275453&-3 & 0 \\
\hline\\
$2l_{S}-2\Phi$ &58.37&-132365&-2&0 \\
\hline\\
$M +2\Phi$ & -264.6& 6093 & -430&0\\
\hline \\
$M -2\Phi+2L_{S}$&46.34& -2491&176&0\\
\hline\\
$2l_{S}+2\Phi$ &1490.35&-1786 &0&0\\
\hline\\
$M -2\Phi$ & 78.86& -1816& 128& 0\\
\hline\\
$-M -2\Phi+2L_{S}$&78.86& 606& -43& 0\\
\hline\\
$2M +2\Phi$ & 1490.35& -348 &47&0 \\
\hline \\
$2M -2\Phi+2L_{S}$& 38.41&-35&5& 0 \\
\hline\\
$2M-2\Phi$&58.37&-14 &2&0 \\
\hline\\
$M +2\Phi+2L_{S}$ & 195.26&-6 &0&0\\
\hline\\
$-M+2\Phi+2L_{S}$&-264.6&-1&0&0 \\
\hline\\
$2M +2\Phi+2L_{S}$ &104.47&0&0& 0 \\

\end{tabular}
}
\end{center}
\end{table}

For Venus, the largest nutation coefficient in longitude for the terms depending on its triaxiality has a 0".6 amplitude  and the largest nutation coefficient
in obliquity has a 0".03 amplitude. Their period corresponds to half the period of rotation of the planets, 121.5 d.

\begin{figure}[htbp]
\center 
\resizebox{.8\hsize}{!}{\includegraphics{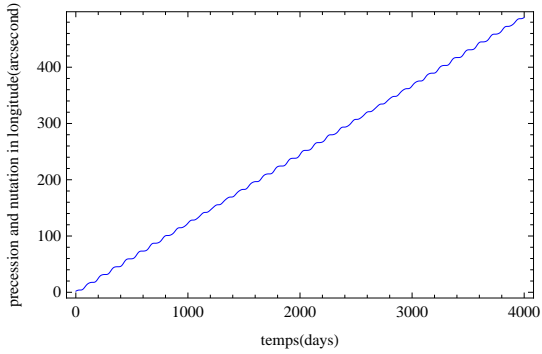}}
\caption{ The precession and the nutation of Venus in longitude for a 4000 day time span.}
 \label{re1}
\end{figure}

\begin{figure}[htbp]
\center 
\resizebox{.8\hsize}{!}{\includegraphics{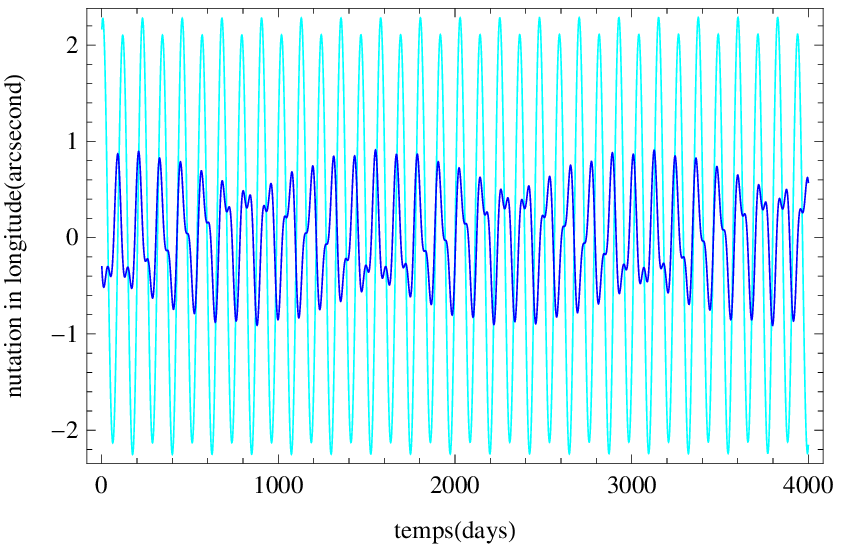}}
\caption{ The nutation in longitude of Venus depending on its dynamical flattening (cyan) and  depending on its triaxiality (blue) for a 4000 day time span, from J2000.0.}
 \label{re4}
\end{figure}

\begin{figure}[htbp]
\center 
\resizebox{.8\hsize}{!}{\includegraphics{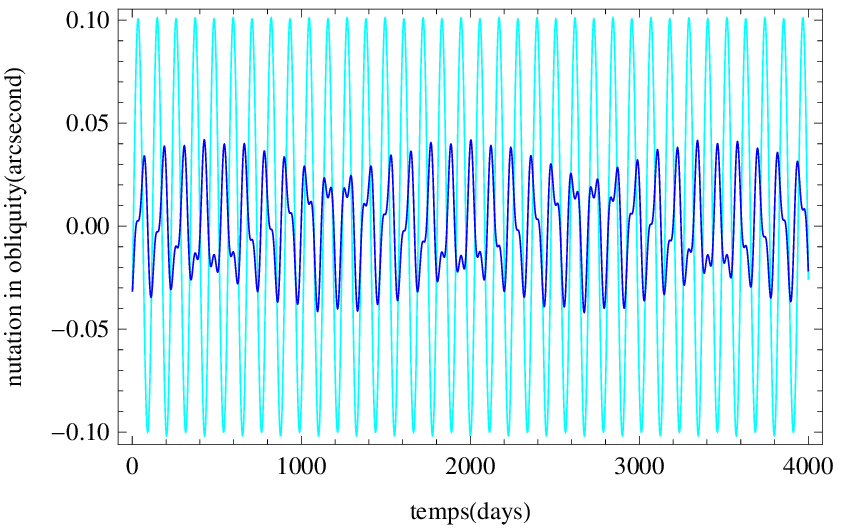}}
 \caption{The nutation in obliquity of Venus depending on its dynamical flattening (cyan) and depending on its triaxiality (blue) for a 4000 day time span, from J2000.0.}
 \label{re5}
\end{figure}

\begin{figure}[htbp]
\center 
\resizebox{.8\hsize}{!}{\includegraphics{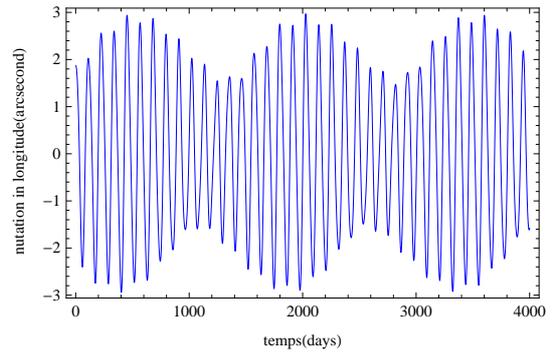}}
 \caption{ The nutation of Venus in longitude for a 4000 day time span, from J2000.0}
 \label{re2}
\end{figure}

\begin{figure}[htbp]
\center
\resizebox{.8\hsize}{!}{\includegraphics{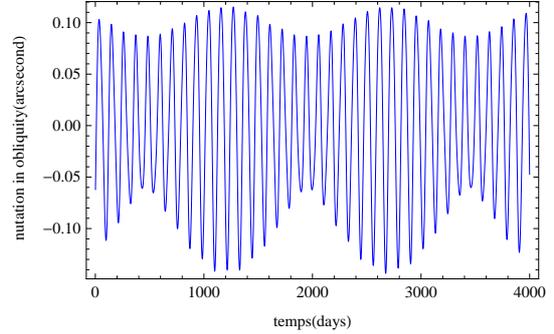}}
 \caption{ The nutation of Venus in obliquity for a 4000 day time span, from J2000.0.}
 \label{re3}
\end{figure}

From the Tables \ref{ww1} to \ref{ww5}, we choose to calculate numerically from their analytical expression the precession-nutation with respect to $t$, over a period of 4000 d. Fig.\ref{re1}. represents the combined precession and nutation of Venus in longitude. The periodic part stands for the nutation and the linear part for the precession (see (\ref{ppp})). In Fig.\ref{re4}. and Fig.\ref{re5}., we represent the nutation of Venus respectively in longitude and in obliquity depending on its dynamical flattening (black curve) and its triaxiality (red curve). In Fig.\ref{re2} and Fig.\ref{re3}., we represent the nutation respectively in longitude and in obliquity when combining the two components above. In Figs.\ref{re1} to \ref{re3}. we choose the 4000 d time span to clearly see the leading oscillations. Recall that our study is valid over 3000 years. The nutation of Venus depending on its triaxiality is dominated by a sinusoid with a period of 121.51 d and another one with a period of 58.37 d. On the other hand the nutation depending on its dynamical flattening is largely dominated by a single sinusoid with a period of 112.35 d. We can also remark (Fig.\ref{re4}. and Fig.\ref{re5}.) that the nutation coefficients of Venus depending on its triaxiality are of the same order of magnitude as the  nutation coefficients depending on its dynamical flattening, whereas for the Earth they are small and negligible in comparison ( Kinoshita,1997 and Souchay et al. 1999). Two reasons explain this : 
\begin{itemize}
\item the ratio of scaling factors is $\frac{K_{a}}{K_{s}}=\frac{1131}{8953}\approx \frac{1}{8}$ for Venus, whereas for the Earth it is two orders of magnitude smaller : $\frac{K_{a}^E}{K_{s}^E}=\frac{5}{3475}\approx \frac{1}{695}$
\item the frequency of the sidereal angle $\dot{\Phi}\approx \dot{l}+\dot{g}$ which enters in the denominator during the integrations in (\ref{riri}) and (\ref{roro}) is roughly 243 times larger for the Earth than for Venus
\end{itemize}.
\\
Thus we have pointed out a significant difference between the nutations of the two planets.

\section{Motion of the pole of Venus in space} \label{9}

From the previous section, it is possible to describe the motion of the pole of Venus with respect to its orbital plane at the date $t$. The rectangular coordinates $(X_{p}, Y_{p}, Z_{p})$ of the pole are given by : 
\begin{eqnarray}\label{mvtpole}
\begin{pmatrix}
X_{p} \\ Y_{p}  \\ Z_{p} 
\end{pmatrix}=
\begin{pmatrix} 
-\sin (I_{0} + \Delta I) \sin (h+\Delta h) \\-\sin (I_{0} + \Delta I) \cos (h+\Delta h)  \\ \cos (I_{0} + \Delta I) \end{pmatrix}
\end{eqnarray}

\begin{figure}[htbp]
\center
\resizebox{0.6\hsize}{!}{\includegraphics{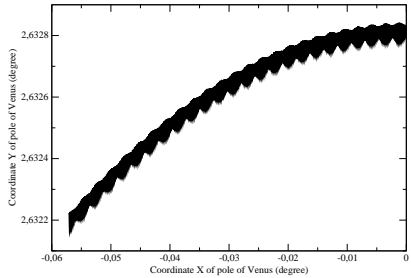}}
 \caption{(X,Y) motion of Venus polar axis in space for a one century time span}
 \label{p1}
\end{figure}

The precession in longitude $\Psi $ is given by
\begin{equation}\label{pre}
\Psi=\dot{\Psi}t=-h=4474".35t \pm 66.5 \ (t \ in \ Julian/cy)
\end{equation}
$I_{0}$ is the nominal value of the obliquity :
\begin{equation}
I_{0}=-\epsilon_{0}=-2.63.
\end{equation}
\
Replacing $\Delta I$ and $\Delta h$ by their numerical values from Tables \ref{ww1} to \ref{ww5}, we show in Fig.\ref{p1}. the motion of the Venus figure axis in space over a one century time interval, for which the combined motion of precession and nutation appears clearly.

From (\ref{pre}) we directly determine the period of precession of Venus, i.e $28965.10 \pm 437 $ years. It is slightly longer than the period of precession of the Earth, i.e. 25712.4 y (Bretagnon et al. 1997).
 Notice that in the case of the Earth, the Moon contributes roughly 2/3 and the Sun 1/3 to the precession rate. In the case of Venus, only the Sun contributes significantly to the precession rate. However, as we have shown in section \ref{8}, this solar contribution is roughly 2.5 times greater than the corresponding one for the Earth. This explains why the periods of precession are rather equivalent.

\section{Numerical values of $\frac{2C-A-B}{2C}$ and $\frac{A-B}{4C}$ and comparison of results} \label{10}

From the various calculations done in the previous sections, we know that the accuracy in the determination of the precession rate $\dot{\Psi}$ and of the coefficients of nutation depends directly on the quality of determination of the dynamical ellipticity $H_{V}=\frac{2C-A-B}{2C}$ and of the triaxiality $T_{V}=\frac{A-B}{4C}$. Therefore for this purpose it is crucial to obtain accurate values of the moments of inertia. The value of ratio $\frac{C}{MR^{2}}$ is inperfectly known. Kozlovskaya (1966) obtained a value in the [0.321;0.360] interval, Shen and Zhang (1988) in the  [0.321;0.350] one and Yoder(1995) in the [0.331;0.341] one. Williams (private communication) gives the following values relative to the differences of the moments of inertia:
\begin{eqnarray}
&&\frac{C-A}{MR^2}=5.519\times 10^{-6},\frac{C-B}{MR^{2}}=3.290\times 10^{-6},\nonumber\\&&\frac{B-A}{MR^{2}}=2.228\times 10^{-6}
\end{eqnarray}
with these values and adopting $\frac{C}{MR^{2}}=0.3360$, we get: 
\begin{eqnarray}
&&\frac{C-A}{C}=1.643\times 10^{-5},\frac{C-B}{C}=9.79\times 10^{-6},\nonumber\\&&\frac{B-A}{C}=6.631\times 10^{-6}.
\end{eqnarray}
All our calculations in this paper have been done with these values.
When studying the rotation of Venus, Habbibulin (1995) has taken $\frac{C}{MR^{2}}=0.340$ and the  values of the Stokes parameters obtained by Williams (1983) :
\begin{eqnarray}
&&\frac{C-A}{C}=2.3324\times 10^{-5},\frac{C-B}{C}=1.2618\times 10^{-5},\nonumber\\&&\frac{B-A}{C}=1.0706\times 10^{-5}.
\end{eqnarray} 

In Table \ref{angel} we compare our results with those obtained by directly using  the values of Habbibulin (1995) for the physical parameters of Venus, thus showing large differences.

\begin{table}[!h]
\caption{Comparison between our results computed with the recent values of the moment of inertia and those obtained with the values of Habbibulin et al.(1995)}
\label{angel}
\begin{center}
\resizebox{1\hsize}{!}{\begin{tabular}[h]{lccc}
\hline\\
 Principal parameter &  Habbibulin(1995) & This study \\
\hline\\
Triaxiality : $\frac{A-B}{4C}$  & $ -2.67*10^{-6}$ &$ -1.66*10^{-6}$  &   \\
\hline\\
 Dyn. flattening : $\frac{2C-A-B}{2C}$&$ 1.70*10^{-5}$ &$ 1.31*10^{-5}$ &  \\
\hline\\
$K_{s}^{V}$ & $-12275".32/cy$ & $-8957".55/cy \pm 133.29$&  \\
\hline\\
 $K_{a}^{V}$ & $1829".34/cy$ & $1133".28/cy \pm 17.09 $&  \\
\hline\\
Precession &$6133".77t$ & $4474".35t \pm 66.5$&  \\
\hline\\
Period of precession& 21156 years& $28965.10\pm 436.99$ years&\\
\hline\\
Largest nutation coefficient & 3".00 & 2".19 &   \\
\end{tabular}
}
\end{center}
\end{table}

Table \ref{angel} shows that the values of the moments of inertia are crucial for our computations.The comparison with the final results computed by Habbibulin (1995) is not possible because in his paper different angles, parametrisation and more complex developments were used. In another study, Zhang (1988) found a precession rate of 7828".6/cy corresponding to a period of 16555 years. Altough no precise information could be extracted from this paper, the difference is probably due to different values of the moments of inertia. What we highlight here is that different values of moment of inertia yield fairly different results for the precession and the nutation as shown in Table \ref{angel}.

\section{Conclusion and  Prospects}
\label{conclusion}
In this paper we investigated by the rotation of Venus. We adopted a well suited theory of the rotation of a rigid body set up by Kinoshita (1977) and developed in full detail by Souchay et al.(1999)  for the Earth. In a first step we defined the reference frames to the various polar axes of the planet (axis of angular momentum, rotation axis, figure axis) and on its moving orbital plane. We have also given in full detail the parametrization of Venus rotation starting from the set of Andoyer canonical variables with respect to the moving orbital plane. In particular we have determined precisely the motion of the orbit of Venus at $t$ with respect to the orbit at J2000.0, through the parameters $\Pi_{1}$ and $\pi_{1}$, for which we have given polynomial expressions. Moreover we have checked the value of the Venus obliquity (263).\\ Then we calculated the disturbing function due to the gravitational interaction with the Sun. Applying Kinoshita's Hamiltonian analytical developments, we calculated the precession constant of Venus, with a precision and an accuracy better than in previous works (Habillulin, 1995, Zhang, 1988). Our value for the precession in longitude is $\dot \psi  = 4474".35t /cy \pm 66.5. $ which is more than two times larger than the corresponding term for the Earth due to the Sun ($\dot \psi  = 1583".99 /cy $) and slightly smaller than the combined effect of the Moon and of the Sun for the Earth ($\dot \psi  = 5000.3 "/cy $). We have shown that the effect of the very small value of the dynamical ellipticity of Venus ($H_{V} = 1.31\times 10^-5$), which should directly lower  the amplitude of its precession and nutation, is more than fully compensated for by its very slow retrograde rotation. Moreover, one of the specificities of Venus is that its  triaxiality ($T_{V} =-1.66\times 10^-6.$)  is of the same order as its dynamical ellipticity (see value above), unlike what  happens for the Earth for which it is considerably smaller.  Consequently, we have performed a full calculation of the coefficients of nutation of Venus due to the Sun and presented the complete tables of nutation in longitude ($\Delta \Psi$) and obliquity ($\Delta \epsilon$) due both to the dynamical ellipticity and the triaxiality.These tables have never been presented in previous works. We think that this work can be a starting point for further studies dealing with Venus rotation, for which it has set up the theoretical foundation (parametrization, equations of motion etc...), such as a study of precession-nutation over a long time scale, the calculation of Oppolzer terms, the effects of the atmosphere.

\section{Appendix}
\label{Annexe}

Demonstration of equation \ref{relat} :\\
\begin{equation*}
\mathtt{da= \cos c \cdot db +\cos B \cdot dc+\sin b  \sin C \cdot dA}. 
\end{equation*}

Derivating the following classical relation :
\begin{equation*}
\mathtt{\cos a = \cos b  \cos c  \\ +\sin b \sin c \cos A } 
\end{equation*}

we obtain: 
\begin{eqnarray} 
\mathtt { -\sin a \cdot da}\nonumber &= & \mathtt{(-\sin b \cos  c +\sin c \cos b \cos A ) \cdot db}\nonumber \\&& + \mathtt{ (-\cos b \sin c +\sin b \cos c \cos A ) \cdot dc} \nonumber\\&& \mathtt{+ (-\sin b \sin c \sin A ) \cdot dA}.\nonumber \\
\end{eqnarray}

Dividing by - $\sin\mathtt{a} $ we have the following equation:
\begin{eqnarray}\label{yoyo}
\mathtt{da}\nonumber &=&\mathtt{\left( \frac{\sin b \cos c -\sin c \cos b \cos A}{\sin a}\right) \cdot db} \nonumber\\&&\mathtt{+\left(\frac{\cos b\sin c -\sin b\cos c\cos A}{\sin a } \right)\cdot dc }\nonumber \\ &&+ \mathtt{\left(\frac{\sin b\sin c \sin A }{\sin a } \right) \cdot dA }.\nonumber \\
\end{eqnarray} 
In the spherical triangle A,B,C the following equations holds :
\begin{equation*}
\mathtt{\frac{\sin c}{\sin C} = \frac{\sin a}{\sin A}\Rightarrow
 \frac{\sin b\sin c\sin A}{\sin a } = \sin b\sin C}.
\end{equation*}
Thus :
\begin{eqnarray*}
\mathtt{\sin a \cos C} &=&\mathtt{\cos c \sin b - \sin b\cos c \cos A} \\&& \mathtt{\Rightarrow \frac{\sin b\cos c-\sin c \cos b \cos A}{\sin a } = \cos C}
\end{eqnarray*}

\begin{eqnarray*}
\mathtt{\sin a \cos B}& =&\mathtt{ \cos b \sin c - \sin c\cos c \cos A }\nonumber\\&&\mathtt{\Rightarrow \frac{\sin c\cos b-\sin b \cos c \cos A}{\sin a } = \cos B}
\end{eqnarray*}
Replacing in  (\ref{yoyo}), we obtain (\ref{relat}).

\end{document}